\documentclass[11pt]{article}
\usepackage{aas_macros,amsmath,amssymb,comment,cite,esint,graphicx,mathtools,diagbox}
\usepackage[margin=.8in,letterpaper]{geometry}
\usepackage[colorlinks=true]{hyperref}
\usepackage[affil-it]{authblk}
\usepackage{subcaption}
\usepackage[utf8]{inputenc}
\usepackage{mathrsfs}
\usepackage{appendix}
\usepackage{amssymb}
\usepackage{float}                  % 图片浮动位置,强制H
\usepackage{color}
\usepackage{cite}
\usepackage{hyperref}
\hypersetup{pageanchor=false}
\usepackage{indentfirst}
\usepackage{url}
\usepackage{xfrac}
\usepackage{caption}
\usepackage[numbers,square,comma,sort&compress,merge]{natbib}
\usepackage{esint}
\usepackage{overpic}
\usepackage{graphicx}
\usepackage{epsf,amsmath,bbold,amsfonts,stmaryrd}
\usepackage{textcomp}
\usepackage{ulem}
\usepackage{tikz}
\usepackage{multirow}

\numberwithin{equation}{section}
\let\originalleft\left
\let\originalright\right
\renewcommand{\left}{\mathopen{}\mathclose\bgroup\originalleft}
\renewcommand{\right}{\aftergroup\egroup\originalright}
\mathcode`\*="8000
{\catcode`\*=\active\gdef*{\mathclose{}\,\mathopen{}}}

\newcommand{\br}[1]{\left[#1\right]}

\newcommand{\pa}[1]{\left(#1\right)}

\newcommand{\be}{\begin{equation}}
\newcommand{\ee}{\end{equation}}
\newcommand{\bea}{\setlength\arraycolsep{2pt} \begin{eqnarray}}
\newcommand{\eea}{\end{eqnarray}}
\newcommand{\nn}{\nonumber}

\newcommand{\M}[1]{\mathcal{#1}}
\newcommand{\isco}{\text{ISCO}}
\newcommand{\df}{\mathrm{d}}
\newcommand{\blue}[1]{\textcolor[RGB]{0,0,255}{ #1}}

\def\D{\Delta}
\def\f{\frac}

\def\nn{\nonumber}

\def\S{\Sigma}
\def\th{\theta}

\def\be{\begin{equation}}
\def\ee{\end{equation}}
\def\bag{\begin{aligned}}
\def\eag{\end{aligned}}
\def\bea{\begin{eqnarray}}
\def\eea{\end{eqnarray}}
\def\ba{\begin{array}}
\def\ea{\end{array}}

\def\bc{\begin{center}}
\def\ec{\end{center}}

\begin{document}
\title{Images and flares of geodesic hotspots around a Kerr black hole}

\author{
Jiewei Huang$^{1}$, Zhenyu Zhang$^{2}$, Minyong Guo$^{1,3\ast}$, Bin Chen$^{2,4,5}$}
\date{}

\maketitle

\vspace{-10mm}

\begin{center}
{\it
$^1$ Department of Physics, Beijing Normal University,
Beijing 100875, P. R. China\\\vspace{4mm}

$^2$Department of Physics, Peking University, No.5 Yiheyuan Rd, Beijing
100871, P.R. China\\\vspace{4mm}

$^3$Key Laboratory of Multiscale Spin Physics, Ministry of Education, Beijing 100875, P. R. China\\\vspace{4mm}

$^4$Center for High Energy Physics, Peking University,
No.5 Yiheyuan Rd, Beijing 100871, P. R. China\\\vspace{4mm}

$^5$ Collaborative Innovation Center of Quantum Matter,
No.5 Yiheyuan Rd, Beijing 100871, P. R. China\\\vspace{2mm}
}
\end{center}

\vspace{8mm}

\begin{abstract}

In this study, we develop a numerical method to generate images on an observer's screen, formed by radiation from hotspots on any timelike orbits outside a black hole. This method uses the calculation of fractional numbers, enabling us not only to produce the overall image but also to distinguish between primary, secondary, and higher-order images. Building upon this, we compute the images of hotspots from eight potential types of geodesic timelike orbits outside a Kerr black hole, summarizing the properties of both the overall and individual order images. Furthermore, we calculate the centroid motion and lightcurve. Notably, we observe flare phenomena across all orbit types and classify these flares into three categories based on the Doppler and gravitational redshift effects.

\end{abstract}

\vfill{\footnotesize $\ast$ Corresponding author: minyongguo@bnu.edu.cn}

\maketitle

\newpage
\baselineskip 18pt
\section{Introduction}\label{sec1}

In recent years, there has been a growing body of observational evidence for the existence of supermassive black holes at the centers of galaxies. Apart from the black hole image captured by the Event Horizon Telescope (EHT) collaboration \cite{EventHorizonTelescope:2019dse,EventHorizonTelescope:2022wkp}, the GRAVITY collaboration has reported intriguing near-infrared flare events originating from near the horizon of the black hole Sgr A* at the center of the Milky Way \cite{GRAVITY:2018det,GRAVITY:2023avo}. While extensive research has been conducted on the black hole image, as referenced in \cite{Gyulchev:2020cvo,Guerrero:2021ues,Hou:2022eev,Qin:2023nog,Meng:2023htc,Guo:2023zwy,Zhang:2024lsf}, our current work will focus on these flare events.

Bright flaring from accreting black holes can be observed across all wavelengths, yet the mechanism driving near-infrared flares from the accretion onto Sgr A* remains a significant point of contention \cite{Ripperda:2021zpn}. The source of these flares is commonly linked to the acceleration of electrons in a localized flaring area that's no larger than a few gravitational radii, rather than a universal increase in the accretion rate or jet power \cite{GRAVITY:2021hxs}. 
There are four main methods to study sources that can produce near-infrared flares: 1) applying general relativistic magnetohydrodynamic simulations to a confined area around the black hole \cite{Dexter:2020cuv,Porth:2020txf,Ripperda:2020bpz,Ripperda:2021zpn,Scepi:2021xgs,Jia:2023iup}, 2) using semi-analytic models to analyze the spectral energy distribution and multi-wavelength variability of Sgr A* \cite{Yuan:2003dc,Younsi:2015xna,ball2021plasmoid,Aimar:2023kzj}, 3) running Particle-in-Cell (PIC) simulations on a small portion of the accretion flow \cite{Rowan:2017cao,Levinson:2018arx,Galishnikova:2022mjg,ElMellah:2023sun,Zhdankin:2023wch}, and 4) performing calculations on hotspots orbiting the central black hole \cite{Broderick:2005my,Meyer:2006fd,Trippe:2006jy,Hamaus:2008yw,Zamaninasab:2009df,Dokuchaev:2020rye,Rosa:2023qcv,Rosa:2024bqv,Chen:2023qic,Chen:2024ilc}. 

Each of the methods mentioned above has its strengths and weaknesses. The first method excels in simulating fluid dynamics around a black hole, but it is computationally demanding and its results do not coincide with observational data. The second method aligns with observational results, but it doesn't adequately explain the cause of flares and the distribution of matter around the black hole. The third method is superior in understanding the microphysical properties of the accretion flow, but its high computational cost and limited calculation area pose challenges for studying sources of realistic sizes. The fourth method, although deemed unable to explain the flare-causing mechanism or the microphysical properties of the accretion flow, has a smaller computational cost. This allows for the study of different black hole parameters and hotspot orbits, theoretically enhancing our understanding of flares \cite{Matsumoto:2020wul,Wielgus:2022heh,vonFellenberg:2023hit}. In previous studies using the fourth method, semi-analytical hotspot imaging models were commonly employed. These models typically use numerical methods to calculate photon orbit trajectories, but often confine the timelike orbits of hotspot motion to analytically solvable, essentially circular, equatorial orbits. Recognizing that there's no inherent necessity for hotspots to remain on the equatorial plane, a significant advancement was made in \cite{GRAVITY:2020lpa}. The introduction of the NERO code signified a development in computing images for any particle orbits outside a Kerr black hole, not just equatorial circular ones. Furthermore, they took into account flares produced by hotspots maintaining a constant vertical component.

Inspired by this pioneering work and in view of a recent classification of Kerr timelike geodesics \cite{Compere:2021bkk}, we aim to develop a numerical method capable of calculating images for any timelike orbits in this work. We intend to compute the images of all possible timelike hotspots outside a Kerr black hole. Specifically, we differentiate primary images, secondary images, and higher-order images by calculating the fractional number. We not only analyze the centroid motion of the complete image but also study the characteristics of different order images. We also calculate the lightcurve to investigate the properties of flares. Particularly, we note that the intensity of the hotspot image is mainly determined by the motion state of the hotspot and gravitational redshift, i.e., the Doppler effect and gravitational redshift jointly dominate the intensity. By analyzing, we ascertain how the Doppler effect and gravitational redshift play roles when each type of orbit generates a flare. 

The structure of this paper is as follows. In Sec. \ref{sec2}, we revisit the classification of timelike geodesics for a Kerr black hole. Sec. \ref{sec3} introduces the hotspot model and imaging methods we employ. In Sec. \ref{sec4}, we apply the model from Sec. \ref{sec3} to generate images from various viewing angles, demonstrating results for all orbit types. We wrap up with a summary of our results in Sec. \ref{sec5}. Throughout this paper, we use the geometrized unit system where $G=c=1$.

\section{Review on classification of radial Kerr geodesic motion}\label{sec2}

In this section, we aim to review the classification of radial Kerr geodesic motion as developed in the paper \cite{Compere:2021bkk}. The line element of the Kerr metric in Boyer-Lindquist (BL) coordinates is presented in the following form
\bea
\mathrm{d}s^2&=&-\left(1-\frac{2Mr}{\Sigma}\right)\mathrm{d}t^2+\frac{\Sigma}{\Delta}\mathrm{d}r^2+\Sigma\mathrm{d}\theta^2+\left(r^2+a^2+\frac{2Mra^2}{\Sigma}\mathrm{sin}^2\theta\right)\mathrm{d}\phi^2-\frac{4Mra}{\Sigma}\mathrm{sin}^2\theta\mathrm{d}t\mathrm{d}\phi\,,\nn\\
\eea
where 
\bea
\Delta=r^2-2Mr+a^2\,,\quad\quad\Sigma=r^2+a^2\cos^2\theta\,.
\eea
From this point forward, for convenience and without loss of generality, we set $M=1$. The outer event horizon of the black hole can then be expressed as:
\be
r_+=1+\sqrt{1-a^2}\,,
\ee
which represents the larger root of the equation $\Delta=0$.

The timelike geodesic equations in Kerr spacetime can be fully integrated due to the presence of four conserved quantities along the trajectory of a particle. These quantities include: the mass\footnote{For the sake of simplifying calculations, we can assume the mass of the timelike object to be unity, without sacrificing generality.} $\mu^2=-g_{\mu\nu}p^\mu p^\nu=1$, the energy $E=-u\cdot \partial_t$, the angular momentum $L=u\cdot\partial_\phi$, and the Carter constant $Q$. By utilizing these conserved quantities, the equations of motion can be expressed as follows \cite{Bardeen:1973tla}
\bea
\label{pr}
u^r&=&\pm_r\f{1}{\S}\sqrt{\M{R}(r)}\,,\nn\\
\label{pth}
u^\theta&=&\pm_\theta\f{1}{\S}\sqrt{\Theta(\theta)}\,,\nn\\
\label{pphi}
u^\phi&=&\f{1}{\S}\br{\f{a}{\D}[E(r^2+a^2)-aL]+\f{L}{\sin^2\theta}-aE}\,,\nn\\
\label{pt}
u^t&=&\f{1}{\S}\br{\f{r^2+a^2}{\D}[E(r^2+a^2)-aL]+a(L-aE\sin^2\theta)}\,,
\eea
where
\bea
\label{Rr}
\M{R}(r)&=&[E(r^2+a^2)-aL]^2-\D[Q+(L-aE)^2+r^2]\,,\\
\label{Thth}
\Theta(\theta)&=&Q+a^2(E^2-1)\cos^2\th-L^2\cot^2\theta\,,
\eea
are the radial and angular potentials respectively, with $\pm_r$ and $\pm_\theta$ denote the signs of the radial and polar motions. The classification of timelike geodesics is based on the root structure of the radial potential $\M R (r)$. We label the out-of-horizon roots of $\M R (r)$ as $r_1$, $r_2$ and $r_3$ where $r_+\leq r_1<r_2<r_3$. urthermore, we adopt the notations from \cite{Compere:2021bkk}: The symbols $\vert$, $+$, $-$ and $\rangle$ denote the outer horizon $r_+$,  a region where motion is permitted ($\M R (r)>0$), a region where motion is disallowed ($\M R (r)<0$) and the radial infinity. The symbols $\bullet$, $\bullet\hspace{-1.5pt}\bullet$, $\bullet\hspace{-4pt}\bullet\hspace{-4pt}\bullet$ and $\vert\hspace{-4.5pt}\bullet$ used to represent a single root, a double root, a triple root, and a root at the outer horizon, respectively.

\begin{table}[h]
\centering
\begin{tabular}{|c|c|c|c|c|}
\hline
Type & Radial range & Root structure & Energy range & $\M{N}$ (of free parameters)\\
\hline
$\M P$ & $r_+\leq r<\infty$ & $\vert+\rangle$ & $E\geq 1$ & $3$\\
\hline
\multirow{2}*{$\M D$} & \multirow{2}*{$r_2\leq r<\infty$} & $\vert+\bullet-\bullet+\rangle$ & $E\geq 1$ & $3$\\
	\cline{3-5}
	& & $\vert\hspace{-6.6pt}\bullet-\bullet+\rangle$ & $E\geq 1$ & $2$\\
\hline
\multirow{4}*{$\M T$} & \multirow{4}*{$r_+\leq r\leq r_1$} & $\vert+\bullet-\rangle$ & $0\leq E<1$ & $3$\\
	\cline{3-5}
	& & $\vert+\bullet-\bullet+\bullet-\rangle$ & $E_{\isco^+}<E<1$ & $3$\\
	\cline{3-5}
	& & $\vert+\bullet-\bullet+\rangle$ & $E\geq 1$ & $3$\\
	\cline{3-5}
	& & $\vert+\bullet-\bullet\hspace{-4pt}\bullet-\rangle$ & $E_{\isco^+}<E<1$ & $2$\\
\hline
\multirow{2}*{$\M B$} & \multirow{2}*{$r_2\leq r\leq r_3$} & $\vert+\bullet-\bullet+\bullet-\rangle$ & $E_{\isco^+}<E<1$ & $3$\\ 
  	\cline{3-5}
	& & $\vert\hspace{-6.6pt}\bullet-\bullet+\bullet-\rangle$ & $E_c<E<1$ & $2$\\
\hline
\multirow{5}*{$\M S$} & \multirow{2}*{$r=r_1=r_2$} & $\vert+\bullet\hspace{-4pt}\bullet+\bullet-\rangle$ & $E_{\isco^+}<E<1$ & $2$\\
	\cline{3-5}
	& & $\vert+\bullet\hspace{-4pt}\bullet+\rangle$ & $E\geq 1$ & $2$\\
	\cline{2-5}
	& \multirow{2}*{$r=r_2=r_3$} & $\vert+\bullet-\bullet\hspace{-4pt}\bullet-\rangle$ & $E_{\isco^+}<E<1$ & $2$\\
	\cline{3-5}
	& & $\vert\hspace{-6.6pt}\bullet-\bullet\hspace{-4pt}\bullet-\rangle$ & $E_c\leq E<1$ & $1$\\
	\cline{2-5}
	& $r=r_1=r_2=r_3$ & $\vert+\bullet\hspace{-4pt}\bullet\hspace{-4pt}\bullet-\rangle$ & $E_{\isco^+}<E<E_{\isco^-}$ & $1$\\
\hline
$\M H$ & $r_1<r\leq r_2$ & $\vert+\bullet\hspace{-4pt}\bullet+\bullet-\rangle$ & $E_{\isco^+}<E<1$ & $2$\\
\hline
$\M {WD}$ & $r_1<r<\infty$ & $\vert+\bullet\hspace{-4pt}\bullet+\rangle$ & $E\geq 1$ & $2$\\
\hline
\multirow{3}*{$\M {WT}$} & \multirow{3}*{$r_+\leq r<r_1$} & $\vert+\bullet\hspace{-4pt}\bullet+\bullet-\rangle$ & $E_{\isco^+}<E<1$ & $2$\\
	\cline{3-5}
	& & $\vert+\bullet\hspace{-4pt}\bullet+\rangle$ & $E\geq 1$ & $2$\\
	\cline{3-5}
	& & $\vert+\bullet\hspace{-4pt}\bullet\hspace{-4pt}\bullet-\rangle$ & $E_{\isco^+}<E<E_{\isco^-}$ & $1$ \\
\hline
\end{tabular}
\caption{Classification of timelike geodesics in the Kerr exterior.}
\label{table:classification}
\end{table}

As indicated in \cite{Compere:2021bkk}, timelike geodesics in the Kerr exterior are categorized into eight classes with $E\ge0$, as outlined in Table \ref{table:classification}. This table can be understood as follows: the first column represents the types of orbits, which include plunging orbits $\M P$, deflecting orbits $\M D$, trapped orbits $\M T$, bounded orbits $\M B$, spherical orbits $\M S$, homoclinic orbits $\M H$, whirling deflecting orbits $\M {WD}$, and whirling trapped orbits $\M {WT}$. Specifically, plunging orbits refer to particles that either plunge into the black hole from infinity or are emitted near the black hole and fly off to infinity. Deflecting orbits describe particles that originate from infinity, reach a turning point, and then rebound back to infinity. Trapped orbits represent particles that are emitted near the black hole, reach a turning point at a finite radius, and subsequently plunge into the black hole. Bounded orbits refer to particles that oscillate between two radial turning points. Furthermore, in spherical orbits, a particle maintains a constant orbital radius $r$. In homoclinic orbits, a particle originates from an unstable spherical orbit, bounces off a turning point, and then approaches the original unstable spherical orbit again. Whirling deflecting particles originate from an unstable spherical orbit and fly off to infinity, or alternatively, they come from infinity and approach an unstable spherical orbit. Lastly, whirling trapped particles either emit outward near the black hole, asymptotically approaching a spherical orbit, or they originate from a spherical orbit before plunging into the black hole. 

The second column of Table \ref{table:classification} is relatively easy to comprehend, as it denotes the range of the corresponding orbits in the radial coordinate. The third column, on the other hand, may be slightly confusing. It represents the complete root structure of the radial function $\M{R}(r)=0$, corresponding to the orbits outside the horizon. In other words, the range of the corresponding orbits in the radial coordinate is a subinterval of the root structure. Other intervals of the root structure may correspond to different types of timelike orbits. The fourth column is also easy to understand; it represents the range of energy that particles on the orbit can carry. The fifth column $\M N$ indicates the number of freely selectable values among the three conserved quantities $E$, $L$, and $Q$ for particles on the corresponding orbit. Given that the expression for  $\M{R}(r)$ includes the three variable parameters $E$, $Q$, and $L$, aside from $r$, these parameters can be freely chosen within permissible ranges when there are no specific restrictions on the roots of $\M{R}(r)$, leading to three degrees of freedom\footnote{For a specific black hole, the spin parameter $a$ is fixed. Furthermore, we have set the mass $\mu=1$. If we retain $\mu$ in the calculations, the free parameters should be scaled as $E/\mu$, $L/\mu$ and $Q/\mu^2$, which is equivalent to choosing $\mu=1$.}. However, when the requirement is for $\M{R}(r)=0$ to have a double root $r_\star$ or a root at $r=r_+$, satisfying $\M{R}^\prime(r_\star)=0$ or $\M{R}(r_+)=0$ would impose a constraint, reducing the number of freely selectable parameters to $2$. When $\M{R}(r)=0$ has a triple root, or a double root with another root precisely on the horizon, the number of free parameters becomes $1$. Specifically, orbits with $\M N=3$ are referred to as generic orbits in \cite{Compere:2021bkk}, while those with $\M N=2$ and $\M N=1$ are respectively termed codimension 1 and codimension 2 orbits. We would like to emphasize that the values of the free parameters are not arbitrary. For a specific type of orbit, they must fall within a certain range.  While the expressions that describe these ranges are quite complex, we will only provide the formulas for some key parameters immediately. For a more detailed understanding, interested readers are encouraged to refer to \cite{Compere:2021bkk} for more information.

For the double root $r_\star$, the conditions $\M{R}(r_{\star} )=\M{R}'(r_{\star} )=0$ hold true. By manipulating these equations, we can also express $Q$ and $L$ as functions of $r_{\star}$ and $E$,
\begin{align}
  Q_{\star}&=
  \dfrac{r_\star^2}{a^2\pa{r_\star-1}^2}
    \bigg( -r_\star^3+3r_\star^2+\pa{a^2-4}r_\star+a^2 \nonumber \\ 
    &+r_\star\pa{1-E^2}\pa{r_\star^3-4r_\star^2+5r_\star-2a^2}
   + 2E\Delta\pa{r_\star}
   \sqrt{r_\star\pa{1+\pa{E^2-1}r_\star}}\bigg)\,,
  \\
  L_{\star}&=\dfrac{  
    E\pa{r_\star^2-a^2}-\Delta\pa{r_\star}
    \sqrt{
      r_\star\pa{1+\pa{E^2-1}r_\star}
    }
  }{a\pa{r_\star-1}}\,.
\end{align}

Take note that in this case $r_\star$ is not fixed and has a lower limit, which is its minimum value. When $r_\star$ reaches this minimum, the corresponding expression for energy $E_c$ can be represented as follows:
\be
\label{Ec}
E_c=\frac{r_+ +2\sqrt{r_+ -1}}{\sqrt{r_+(r_+ + 4\sqrt{r_+ -1}+2)}}\,.
\ee

On the other hand, if the orbit $r=r_\star$ is confined to the equatorial plane, we have $Q_{\star}=0$. In this case, the particle follows a circular trajectory. The corresponding energy can be calculated and expressed as
\be
E_\pm(r_\star)=\frac{r_\star(r_\star-2)\pm ar_\star^{1/2}}{\sqrt{r_\star^3(r_\star-3)\pm2ar_\star^{5/2}}}\,,
\ee
where the symbols $\pm$ represent whether the particle is moving in a prograde or retrograde direction. If we further require the particle to move in the innermost stable circular orbit (ISCO), i.e., where $\M{R}^{\prime\prime}(r)=0$ is satisfied, then the radius of the ISCO can be determined as 
\be
r_{\isco^\pm}=3+Z_2\mp\sqrt{(3-Z_1)(3+Z_1+2Z_2)}\,,
\ee
where
\be
Z_1=1+(1-a^2)^{1/3}\br{(1+a)^{1/3}+(1-a)^{1/3}}\,,\qquad
Z_2=\sqrt{3a^2+Z_1^2}\,.
\ee
Similarly, the symbols $\pm$ here represent prograde and retrograde motion, respectively. Consequently, the energy for the corresponding prograde and retrograde orbits is given by
\bea
\label{Ep}
E_{\isco^+}=E_+(r_{\isco^+})\,,\\
\label{Em}
E_{\isco^-}=E_-(r_{\isco^-})\,.
\eea
By now, we have provided clear explanations for all the parameters mentioned in Table \ref{table:classification}. One should have a sufficient understanding of the orbit classifications and the corresponding energy conditions outlined in the table. We will also provide examples of each class later in Sec. \ref{sec4}.

\section{Hotspot model and imaging method}\label{sec3}

In this section, we introduce our hotspot model and the associated imaging method. Our work primarily focuses on the kinematic flare phenomena brought about by gravitational redshift (blueshift) and Doppler shift. As such, we dismiss relativistic magnetohydrodynamic effects. We represent the hotspot as an opaque object traversing various types of geodesic paths outside a Kerr black hole. For ease of computation, we model the shape of the hotspot as a sphere with a small, fixed radius $b$ in the BL coordinates. Since the minimum event horizon radius of the Kerr black hole can be $1$, the size of the hotspot should be smaller than the scale of the black hole's event horizon. Therefore, for convenience, we will designate $b=0.25$. We do not aim to investigate the influence of the radiation mechanism on imaging and flares. Therefore, our model does not account for the detailed  radiation spectrum of the hotspot’s emission. We thus assume that the hotspot's emission is isotropic and frequency-independent, implying a broadband source with a flat spectrum.

To image the moving hotspot, we need to understand both the trajectory of the light source and the radiation transfer from the source. We determine the trajectory of the source by numerically solving the geodesic equation in the Hamilton-Jacobi form. For the radiation transfer, we employ the numerical backward ray-tracing method and the fisheye camera mode for imaging. The specific technical details can be found in Appendix B of \cite{Hu:2020usx}. The camera is set as a zero-angular-momentum observer (ZAMO), whose tetrads takes the following form
\bea
\hat{e}_{(0)}=\frac{g_{\phi\phi}\partial_t-g_{\phi t}\partial_{\phi}}{\sqrt{g_{\phi\phi}\left(g_{\phi t}^2-g_{\phi\phi}g_{tt}\right)}}\,,\quad\quad \hat{e}_{(1)}=-\frac{\partial_r}{\sqrt{g_{rr}}}\,,\quad
\hat{e}_{(2)}=\frac{\partial_\theta}{\sqrt{g_{\theta\theta}}}\,,\quad \hat{e}_{(3)}=-\frac{\partial_\phi}{\sqrt{g_{\phi\phi}}}\,.
\eea

We consider that the intensity of light emitted from the source and reaching the observer is not absorbed by the medium during its journey. The intensity divided by the frequency cubed, denoted as $I_\nu/\nu^3$, is conserved along a light ray \cite{Lindquist:1966igj}. Thus, the observed and emitted specific intensities $I_{\nu_o}$ and $I_{\nu_s}$ are related by radiative transport as
\be
\frac{I_{\nu_o}}{\nu_o^3}=\frac{I_{\nu_s}}{\nu_s^3}\,,
\label{Inu}
\ee
where $\nu_o=-k_\mu \hat{e}_{(0)}^\mu$0 is the observed frequency on the screen, with $k_\mu$ representing the 4-momentum of the photon. Meanwhile, $\nu_s=-k_\mu u^\mu$ is the frequency measured by an observer comoving with the source. After introducing the redshift factor $g=\nu_o/\nu_s$, Eq. (\ref{Inu}) can be rewritten as
\be
I_{\nu_o}=g^3 I_{\nu_s}\,.
\ee
Note that in our hotspot model, $I_{\nu_s}$ is a constant. Therefore, the observed intensity is solely dependent on the redshift factor, $g$.

\begin{figure}[h]
  \centering
  \includegraphics[width=5.8in]{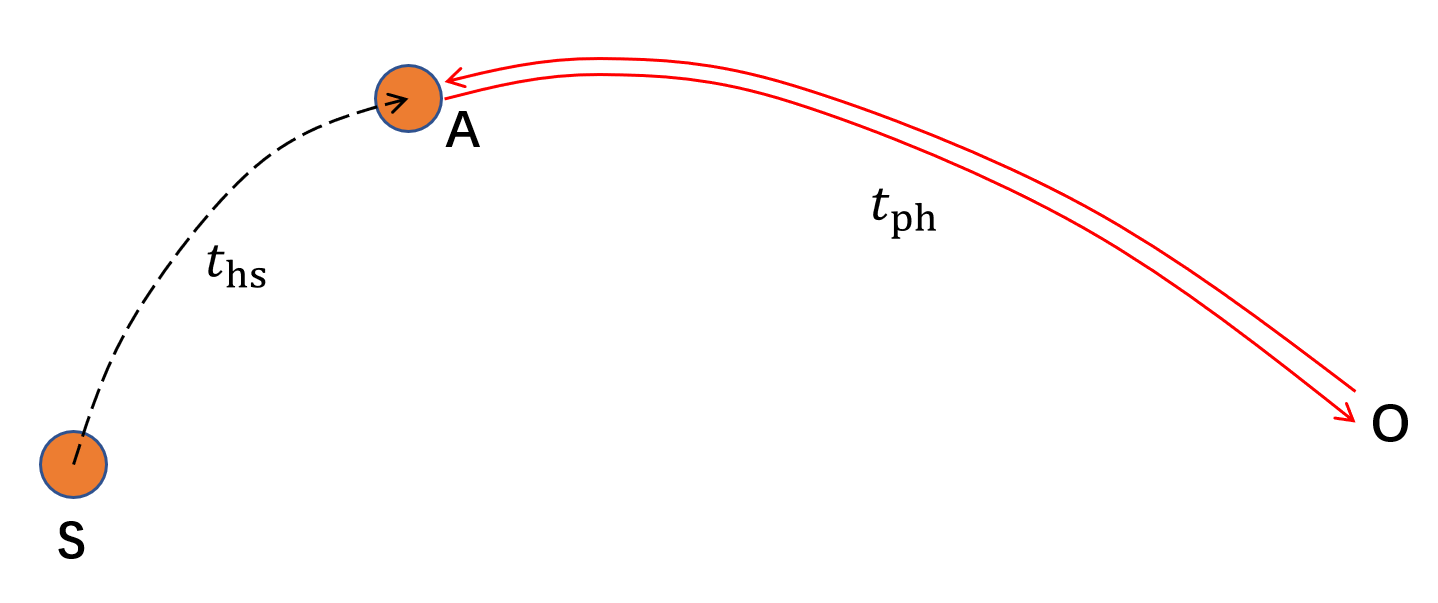}
  \centering
  \caption{A diagram to illustrate the method to calculate the time}
  \label{fig:calt}
\end{figure}

It is important to highlight a subtle issue that arises when handling the imaging of a moving hotspot. The light that is emitted from the hotspot and reaches the observer is not an instant reflection of the hotspot's position. Given that the hotspot is in motion, it will have moved to a new location by the time the light reaches the observer, resulting in a delayed image. Given the substantial distance between the observer and the hotspot, this time delay is significant and cannot be overlooked. To tackle this issue, we have implemented a specific strategy, illustrated in Fig. \ref{fig:calt}, which is consistent with the one presented in the paper \cite{GRAVITY:2020lpa}. Let S represent the starting point of the hotspot and O denote the observer. As the hotspot moves to point A and intersects with the photon traced backward from the observer, we define the coordinate time for the hotspot's journey from S to A as $t_\text{hs}$, and the coordinate time for the photon's journey from O to A as $t_\text{ph}$, all within the BL coordinate system. The total time of interest to us is simply the sum of these two times, $t_\text{hs}+t_\text{ph}$. It is crucial to note that we assume the observer's proper time is equivalent to the coordinate time, rendering this method effective only for observers sufficiently distant from the black hole. 

In addition, due to the strong gravitational lensing effects caused by black holes, the imaging of the hotspot yields not just a primary image, but also secondary and higher-order images. As a result, it becomes necessary for us to distinguish between these different levels of images. To achieve this, we employ a specific method that involves calculating two angular integrals over the entire light path, stretching from the hotspot to the observer. The definitions of these integrals are
\bea
G_\theta=\fint_{\theta_s}^{\theta_o} \dfrac{1}{\pm_\theta 
    \sqrt{\Theta(\theta)}}
    \df \theta\,,\quad\quad \hat{G}_\theta=\int_{\theta_{-}}^{\theta_{+}}
 \dfrac{1}{\sqrt\Theta}\df \theta\,.    
\eea
The integral symbol $\displaystyle \fint$ denotes the integration along the trajectory of a photon trajectory, while $\theta_o$ and $\theta_s$ are the angular coordinates of the observer and the hotspot, respectively. Furthermore, $\theta_\pm$ represents the larger and smaller angular turning points, respectively. Next, we calculate the fractional number $n$ using the formula
\bea
n=\dfrac{G_\theta}{2\hat{G}_\theta}\,.
\eea
When $n\le1/2$, it corresponds to the primary image. If $1/2<n\le1$, it corresponds to the secondary image, and for $n>1$, it corresponds to higher-order images.

Furthermore, despite the assumption that the emission rate of the hotspot is constant within its local coordinate system, due to gravitational redshift (blueshift) and the Doppler effect, the intensity of the image seen by the observer varies at different times. To present the images more effectively, we utilize the definition of $T$ as proposed in \cite{Zhang:2023cuw}:
\bea
T=\left[
{\log_2\left(1+\dfrac{I_o^\text{max}}{I_o}\right)}\right]^{-1}\,,
\label{normI}
\eea
This definition is based on the expression for the brightness temperature \cite{rybicki1991radiative}. Here, $I_o^\text{max}$ is defined as the maximum value of $I_o$, ensuring that $T$ ranges between $0$ and $1$. Additionally, at each moment, we can compute the centroid position $\vec{x}_c$ of the flux across the camera plane. Given the definition of the camera we have adopted \cite{Hu:2020usx}, the flux $F(i,\,j)$ of pixels $(i,\,j)$ observed at the screen can be calculated by
\begin{align}
  F(i,j)=I_o S_0\cos 
  \left[2 \arctan 
  \left( \dfrac{1}{N}
   {\tan \left(\frac{\alpha_\text{fov}}{2}
    \right)}
     \sqrt{\left(i-\frac{{N}+1}{2}\right)^2+
     \left(j-\frac{{N}+1}{2}\right)^2}
     \right)
     \right] \,,
\end{align}
where $S_0$ is the pixel size, \(N\) is the number of horizontal or vertical pixels, \(i\) and \(j\) are chosen in a range from \(1\) to \(N\), and $\alpha_\text{fov}$ is the camera's field of view. Once we have this information, the centroid position for each image can be calculated using the formula:
\begin{align}
   \vec{x}_c(t)=\dfrac{\sum_{i,j}\vec{x}\pa{i,j}
   F\pa{i,j}}{
    \sum_{i,j} F\pa{i,j}
   }\,,
\end{align}
where $\vec{x}\pa{i,j}$ represent the coordinates of pixel \(\pa{i,j}\), and the total flux of the snapshot, denoted as $\sum_{i,j} F\pa{i,j}$ can be regarged as the flux for $\vec{x}_c(t)$. The images we obtained of the moving hotspot are a series of snapshots. To conveniently identify the brighter snapshots, i.e., flares, we introduce the rescaled flux of $ \vec{x}_c(t)$ as
\bea
\bar{F}(\vec{x}_c)=\frac{\sum_{i,j} F\pa{i,j}}{\left[\sum_{i,j} F\pa{i,j}\right]_{\text{max}}}\,.
\eea
In the context of our model and the methodology we've implemented, we ultimately determine the time evolution of both the centroid position and its corresponding flux.

\section{Results}\label{sec4}

In this section, we will display images of a hotspot orbiting around the Kerr black hole with $a=0.94$, utilizing the orbit classifications discussed in the previous section. Considering that the black hole (BH) is located a significant distance from us in the universe, we set $r_o = 300 \gg r_+$ in our numerical calculations. We are specifically examining two scenarios for observational angles: one with $\theta_o = 25 ^\circ$ and another with $\theta_o = 80^\circ$. Initially, we will investigate the characteristics of hotspot imaging at the observational angle of $\theta_o=25^\circ$. Afterwards, we will shift our focus to the observations made at $\theta_o=80^\circ$. In the following calculations, we set the azimuthal angle $\phi_o=0^\circ$.

\begin{table}[H]
  \centering
  \begin{tabular}{|c|c|c|c|c|c|c|}
  \hline
  Type   & $E$ & $L$
  &$Q$& \(r_s\)
  &Root structure
  &Root
  \\
  \hline
  $\M P$ & $1.2$ & $1$ & $12$ & $40$
  &$\blue{\vert+\rangle}$
  & $\diagdown$
  \\ \hline
  $\M D$ & $1.2$ & $3$ & $5$ & $40$  
  &$\vert+\bullet-\blue{\bullet+\rangle}$
  &${1.38,\,\blue{2.38}}$
  \\ \hline
  $\M T$ & $0.95$ & $-1$ & $5$ & $15$
  &$\blue{\vert+\bullet}-\rangle$
  &$\blue{17.51}$
  \\ \hline
  $\M B$ & $0.95$ & $2$ & $5$ & $14$
  &$\vert+\bullet-\blue{\bullet+\bullet}-\rangle$
  &$1.53,\,\blue{3.66},\,\blue{14.78} $
  \\ \hline
  $\M S$ & $0.95$ & $-0.743$ & $12.82$ & $8$
  &$\vert+\bullet-\blue{\bullet}\hspace{-1.5pt}\blue{\bullet}-\rangle$
  &$4.07,\,\blue{8},\,\blue{8}$
  \\ \hline
  $\M H$ & $0.95$ & $0.172$ & $11.86$ & $4.05$
  &$\vert+\blue{\bullet}\hspace{-1.5pt}\blue{\bullet+\bullet}-\rangle$
  &$\blue{4},\,\blue{4},\,11.95$ \\
  \hline
  $\M W\M D$ & $1.2$ & $-4.035$ & $22.47$ & $4.01$
  &$\vert+\blue{\bullet}\hspace{-1.5pt}\blue{\bullet+\rangle}$
  &$\blue{4},\,\blue{4}$
  \\ \hline
  $\M W \M T$ & $0.95$ & $-1.256$ & $12.10$ & $5.99$
  &$\blue{\vert+\bullet}\hspace{-1.5pt}\blue{\bullet}+\bullet-\rangle$
  &$\blue{6},\,\blue{6},\, 8.14$
  \\ 
  \hline
  \end{tabular}
  \caption{Parameters of different types of timelike geodesics.}
  \label{table:parameter of hotspot}
  \end{table}
 
For simplicity, we standardize the initial position to $(t_s,\,\theta_s,\,\phi_s)=(0,\,60^\circ,\,0)$ when selecting parameters for each orbit type. In addition, we use $r_s$ to signify the radial coordinate at which the hotspot commences, and its selection must adhere to the radial range specified for the corresponding orbit in Table \ref{table:classification}. Suitable values for $r_s$ for each type of orbit are given in Table \ref{table:parameter of hotspot}. We opt for $\theta_s=60^\circ$ because our established observation angles fall below $60^\circ$ and above $80^\circ$. As a result, at the initial moment, the images of the hotspot will manifest in the northern and southern hemispheres of the image plane respectively, providing a highly representative view. Additionally, we set $\phi_s=\phi_o=0^\circ$ to position the hotspot between the observer and the black hole at the onset. We must underscore that different types of orbits often correspond to more than a single root structure. Indeed, as illustrated in Table \ref{table:classification}, only the $\M P$, $\M H$ and $\M{WD}$ orbits are associated with a singular root structure, while all other orbit types are not. However, as presented in Table \ref{table:parameter of hotspot}, for each orbit type, we have selectively chosen one parameter set for in-depth study.  Our strategy for selecting parameters involves opting for the scenario where $\M N$ is the largest for a given orbit type. This is because scenarios with smaller $\M N$ values correspond to orbits with repeated roots or roots that fall on the horizon in a peculiar manner. Conversely, scenarios with the largest $\M N$ correspond to more common orbits. Additionally, for the $\M T$, $\M S$, and $\M{WT}$ orbits, the root structure is still not unique when $\M N$ is at its maximum. For the $\M T$ orbit, given that its characteristic is being confined between $r_+$ and $r_1$, we would like to choose orbits with $E<1$. Considering that the $\vert+\bullet-\bullet+\bullet-\rangle$ orbit requires more stringent energy conditions than the $\vert+\bullet-\rangle$ orbit, we opt for the latter. For the S orbit, we naturally desire this orbit to be a stable one, so we select the $\vert+\bullet-\bullet\hspace{-4pt}\bullet-\rangle$ orbit. For the $\M WT$ orbit, similar to the $\M T$ orbit and considering that this orbit is confined between $r_+$ and $r_1$, we choose the case with $E<1$, i.e., the $\vert+\bullet\hspace{-4pt}\bullet+\bullet-\rangle$ orbit. In summary, the eight types of orbits we select can be divided into two categories based on the value of $\M N$. The first category includes $\M P$, $\M D$, $\M T$, and $\M B$ orbits, where $\M N=3$. The second category includes $\M S$, $\M H$, $\M{WD}$, and $\M{WT}$ orbits, with $\M N=2$. For orbits that can extend to infinite distance, such as $\M{P},\M{D},\M{WD}$, we set the energy $E=1.2$ for these specific orbit types. For all other orbit types, we fix the energy at $E=0.95$. Additional specific parameters can be referenced in Table \ref{table:parameter of hotspot}. Notably, within the `root structure' column, regions of the orbit's existence are marked in blue. In the `root' column, the black and blue designations align with those in the `root structure' column. 

\subsection{$\M{N}=3$}

In this subsection, we aim to investigate the imaging characteristics of orbits where $\M N=3$. 

\subsubsection{Plunging orbits}

\begin{figure}[!h]
  \centering
  \includegraphics[width=6in]{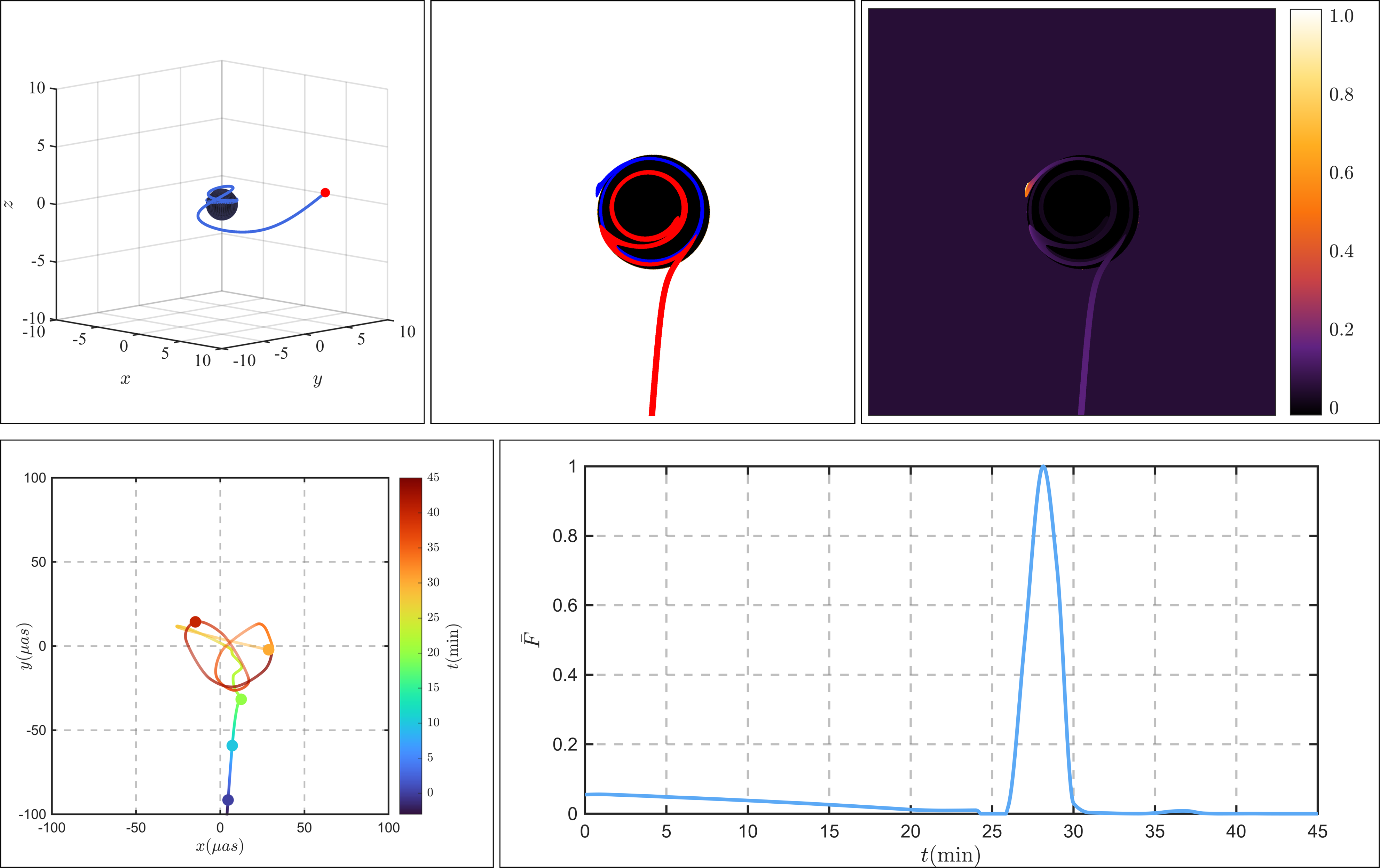}
  \centering
  \caption{The upper row presents the trajectory and image of a plunging orbit, while the lower row illustrates the centroid motion and the light curve of the plunging particle. On the upper row, from left to right, we display: (Left) The trajectory in 3D Cartesian coordinates, defined as $x=r\sin\theta\cos\phi$, $y=r\sin\theta\sin\phi$, $z=r\cos\theta$. The red dot represents the starting point of the trajectory. (Middle) The image of the hotspot as observed by an observer. In this representation, the direct, second, and higher order images are color-coded as red, blue, and orange, respectively. (Right) The intensity map of the image, which offers a visualization of the distribution of light intensity across the image. For the lower row: (Left) The panel depicts the motion of the emission centroid. The accompanying color bar indicates the time progression in units of minutes. The time interval between two adjacent dots in the diagram is 10 minutes and the first dot is 0 min. (Right) The panel presents the flux evolution corresponding to the motion, providing a visual representation of how the rescaled flux changes over time.}
  \label{fig:P}
\end{figure}

We initiate our study with Plunging orbits, denoted as $\M P$. In Fig. \ref{fig:P}, the upper panel displays the trajectory and image of a plunging orbit, while the lower panel illustrates the centroid motion and the light curve for a plunging particle. This scenario represents a particle plunging into the black hole from infinity, observed at an inclination angle of $\theta_o=25^\circ$.

In the left plot of the upper panel of Fig. \ref{fig:P}, we observe a hotspot beginning its journey from a position distant from the black hole and spiraling towards it. Initially, when the hotspot is still far from the black hole, its orbital speed is relatively slow. However, as it gets closer to the black hole, its rotational speed dramatically increases. The corresponding image of this orbit is illustrated in the middle plot of the upper panel. The primary image, represented by the red line, shows the hotspot entering from the southern hemisphere, spiraling, and eventually plunging into the black hole. Additionally, the secondary image, marked by the blue line, is distinctly visible in the northern hemisphere, while the higher order images, indicated by the orange line, are barely noticeable around the shadow curve. The corresponding intensities of these images are displayed in the right plot of the upper panel. For the primary image, we note that as the hotspot moves away from the observer and towards the black hole, its speed increases due to the black hole's gravitational attraction. Consequently, the brightness of the hotspot's image gradually decreases due to the Doppler redshift effect. As it nears the black hole, the brightness of the spiraling image also decreases due to the strong gravitational redshift effect. In contrast, for the secondary image, the light is emitted in a direction roughly parallel to the hotspot's motion as it moves away from the observer and approaches very close to the black hole. After circumnavigating the black hole due to the gravitational lensing effect, the light finally reaches the observer. This scenario, due to the Doppler blueshift effect, results in regions of significantly enhanced brightness.

In the lower panel of Fig. \ref{fig:P}, we observe that the trajectory of the hotspot persists for more than 40 minutes on the observer's screen. In the initial phase, when the hotspot is positioned far from the black hole, the centroid is predominantly determined by the primary image. However, as the hotspot approaches the black hole, roughly 20 minutes into the observation, both the primary and secondary images contribute to determining the centroid. At approximately 28 minutes, there is a noticeable flare, characterized by a sudden surge in the rescaled flux. Comparing this with the results from the upper panel, we can ascertain that this flare is triggered by the Doppler blueshift of the secondary image. Note that although the hotspot emits the light forming the flare right at the onset of its motion, due to the lensing effect, a substantial delay occurs before the flare manifests on the screen. For ease of reference, we term this type of flare as a ``Lensed Doppler Blueshift Flare'' (LDBF).

\subsubsection{Deflecting orbits}

\begin{figure}[!h]
  \centering
  \includegraphics[width=6in]{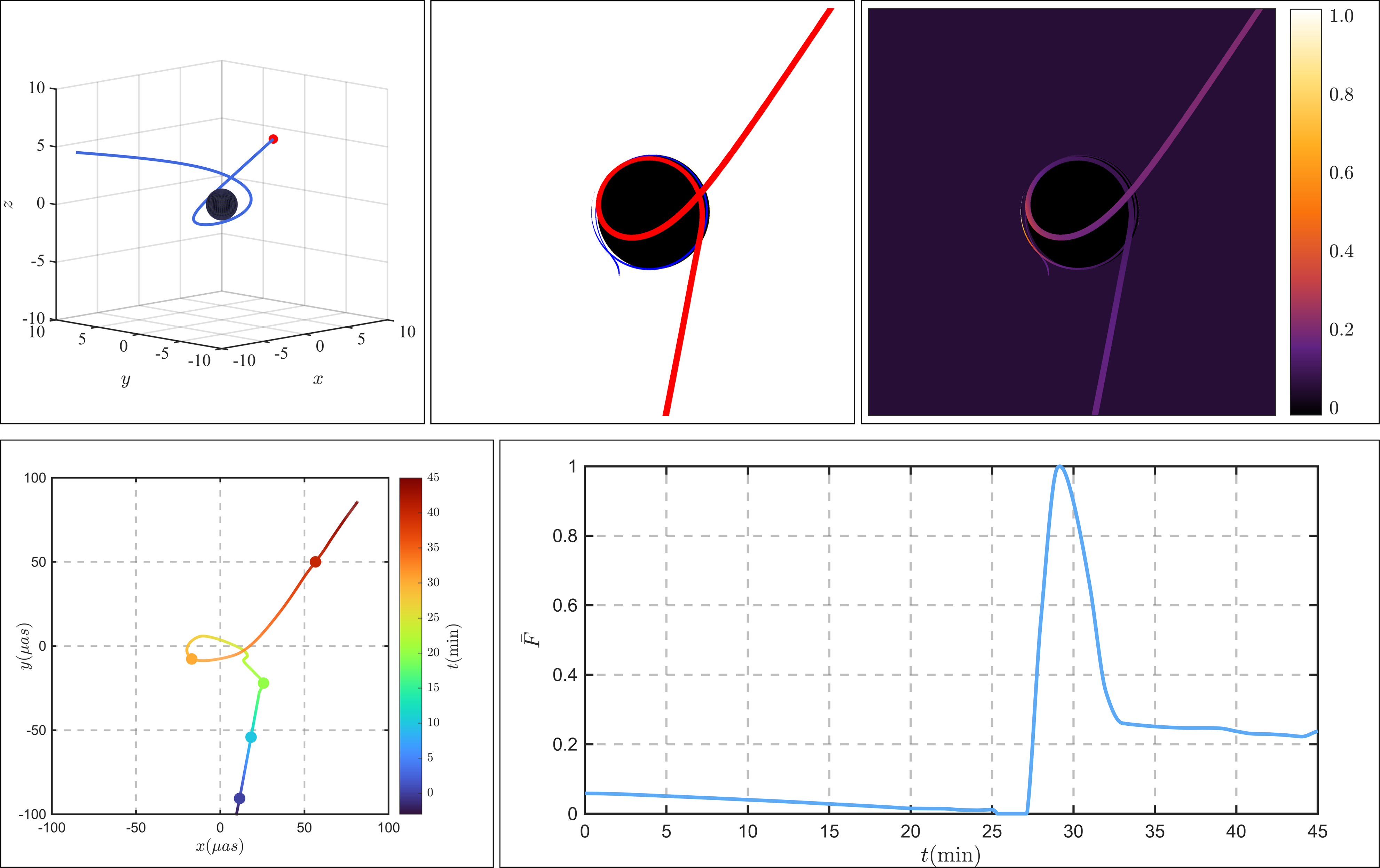}
  \centering
  \caption{The upper row in the illustration presents the trajectory and image of a deflecting hotspot. Conversely, the lower row provides a depiction of the centroid motion and its corresponding light curve for the deflecting orbit. The interpretation of the points and lines in this figure aligns with those outlined in Fig. \ref{fig:P}.}
  \label{fig:D}
\end{figure}

Next, we turn our attention to the study of deflecting orbits, denoted as $\M D$. In Fig. \ref{fig:D}, similar to Fig. \ref{fig:P}, the upper row displays the trajectory and image of a deflecting hotspot. The lower row, on the other hand, delineates the motion of the emission centroid and its associated light curve for the deflecting orbit. The observation is made at an inclination angle of $\theta_o=25^\circ$.

From Fig. \ref{fig:D}, it is observed that the hotspot initiates its journey from a position far from the black hole, reaches a turning point, and then returns to infinity. On the observer's screen, the primary image of the hotspot departs from the northern hemisphere towards the black shadow, circles within the shadow, then flies into the southern hemisphere. The secondary images are primarily distributed in the northern hemisphere near the shadow curve, while the higher-order images clinging to the shadow curve are quite inconspicuous. From the bottom-left graph, it can be deduced that the centroid motion is determined by the primary image. The bottom-right graph reveals that the light curve can be essentially divided into three segments. In the first segment, the flux of the centroid decreases slightly. The second segment presents a distinct flare, while the third segment is relatively stable, yet with a higher flux than at the start. This phenomenon can be explained by the Doppler effect. During the first segment, the hotspot is moving away from the observer, causing a Doppler redshift. In contrast, during the third segment, the hotspot is moving towards the observer, causing a Doppler blueshift. However, as it is not directly facing the observer, the flux is only slightly higher than in the first segment. The appearance of the flare is due to the hotspot passing the turning point at this time. Its speed rapidly increases, and due to its proximity to the black hole, the strong gravitational force can cause a significant change in direction. Therefore, there is a moment when the speed is directly or nearly directly facing the observer, causing a Doppler blueshift that appears as a flare. In this paper, we refer to this type of flare as a `turning Doppler blueshift flare' (TDBF).

\subsubsection{Trapped orbits}

\begin{figure}[!h]
  \centering
  \includegraphics[width=6in]{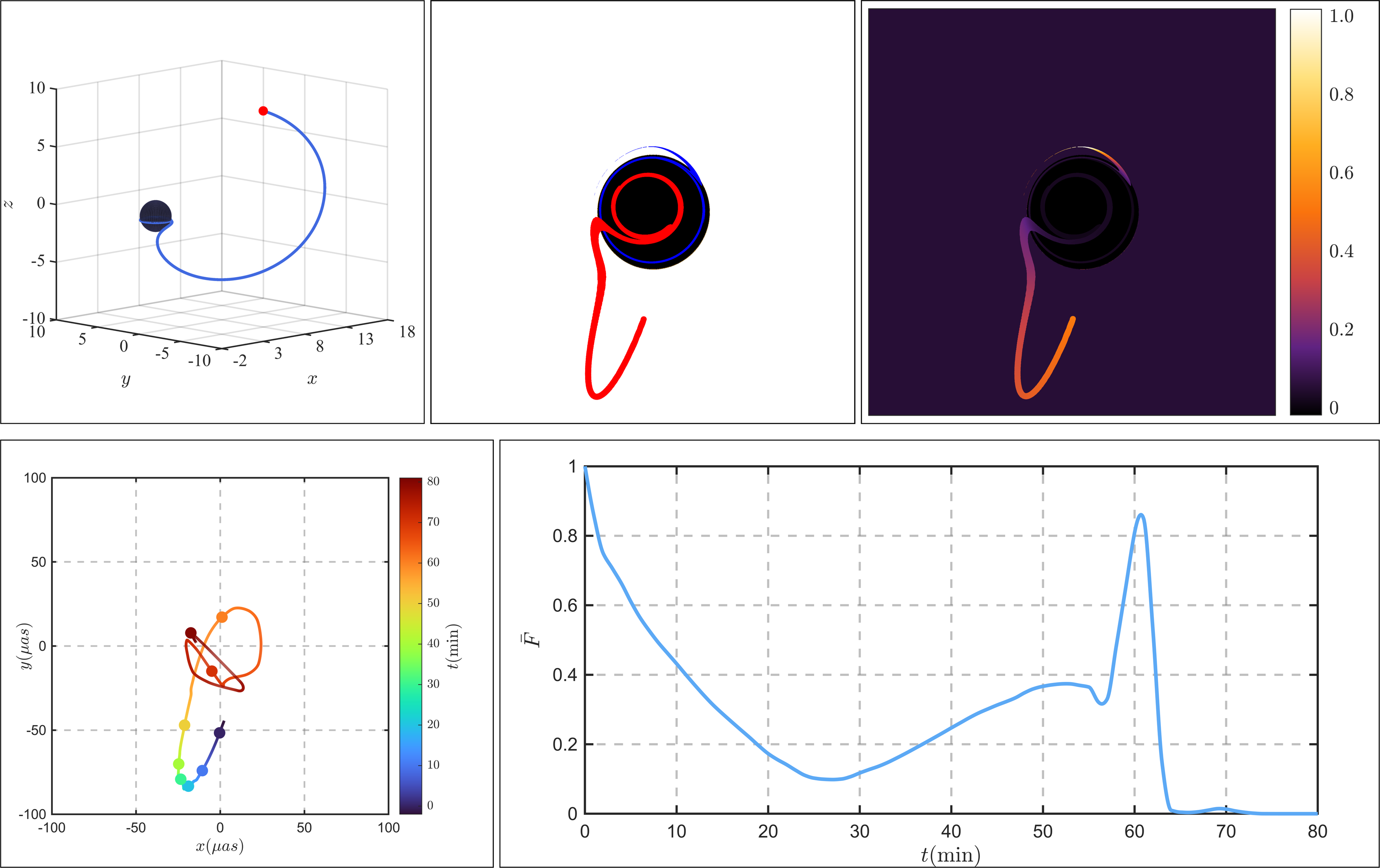}
  \centering
  \caption{The upper row of the illustration showcases the trajectory and image of a hotspot trapped in orbit. In contrast, the lower row visually represents the centroid motion and the corresponding light curve for this trapped orbit. The symbolism and interpretation of points and lines in this figure are consistent with those defined in Fig. \ref{fig:P}.}
  \label{fig:T}
\end{figure}

Then, we will turn our attention to the exploration of trapped orbits, denoted as $\M T$. In Fig. \ref{fig:T}, akin to Fig. \ref{fig:P}, the upper row illustrates the trajectory and image of a hotspot trapped in orbit. Conversely, the lower row outlines the movement of the emission centroid, along with its corresponding light curve for this trapped orbit. All observations are made at an inclination angle of $\theta_o=25^\circ$.

From Fig. \ref{fig:T}, we observe that the hotspot embarks on its journey from $r_s$, subsequently moving outward to reach $r_1$. After this, it changes direction, circles the black hole several times, and ultimately falls into the black hole. On the observer's screen, the primary image of the hotspot starts in the southern hemisphere, then moves downward. After reaching a turning point, it changes direction and finally spirals into the black hole's shadow. The starting point of the secondary image, marked in blue, is near the equatorial plane and close to the black hole's shadow, and then it spirals and falls into the black hole's shadow. From the light intensity plot at the top right and the light curve graph at the bottom right, we can see that the flux of the hotspot's image is highest when it starts moving, then decreases, rises again after more than 20 minutes, and presents a noticeable flare at 60 minutes. The reason for this phenomenon is that initially, the hotspot is moving away from both the black hole and the observer, and the angle between the direction of the hotspot's velocity and the direction pointing towards the observer is increasing, leading to a gradual decrease in light intensity. As the hotspot is captured and flies towards the black hole, the angle between the direction of the hotspot's velocity and the direction pointing towards the observer decreases. During this process, the hotspot is far from the black hole, so the effect of gravitational redshift is small, and the received light intensity gradually increases. As the hotspot nears the black hole, despite the increasing gravitational redshift, the impact of the Doppler effect on the secondary image exceeds the effect of gravitational redshift, resulting in a flare. This flare is similar to those observed in plunging orbits and is characterized as a LDBF. Correspondingly, the centroid motion graph at the bottom left can also be understood. Initially, the centroid is dominated by the primary image. Subsequently, it is dominated by the secondary image, and finally, it returns to being dominated by the primary image.

\subsubsection{Bounded orbits}

\begin{figure}[!h]
  \centering
  \includegraphics[width=6in]{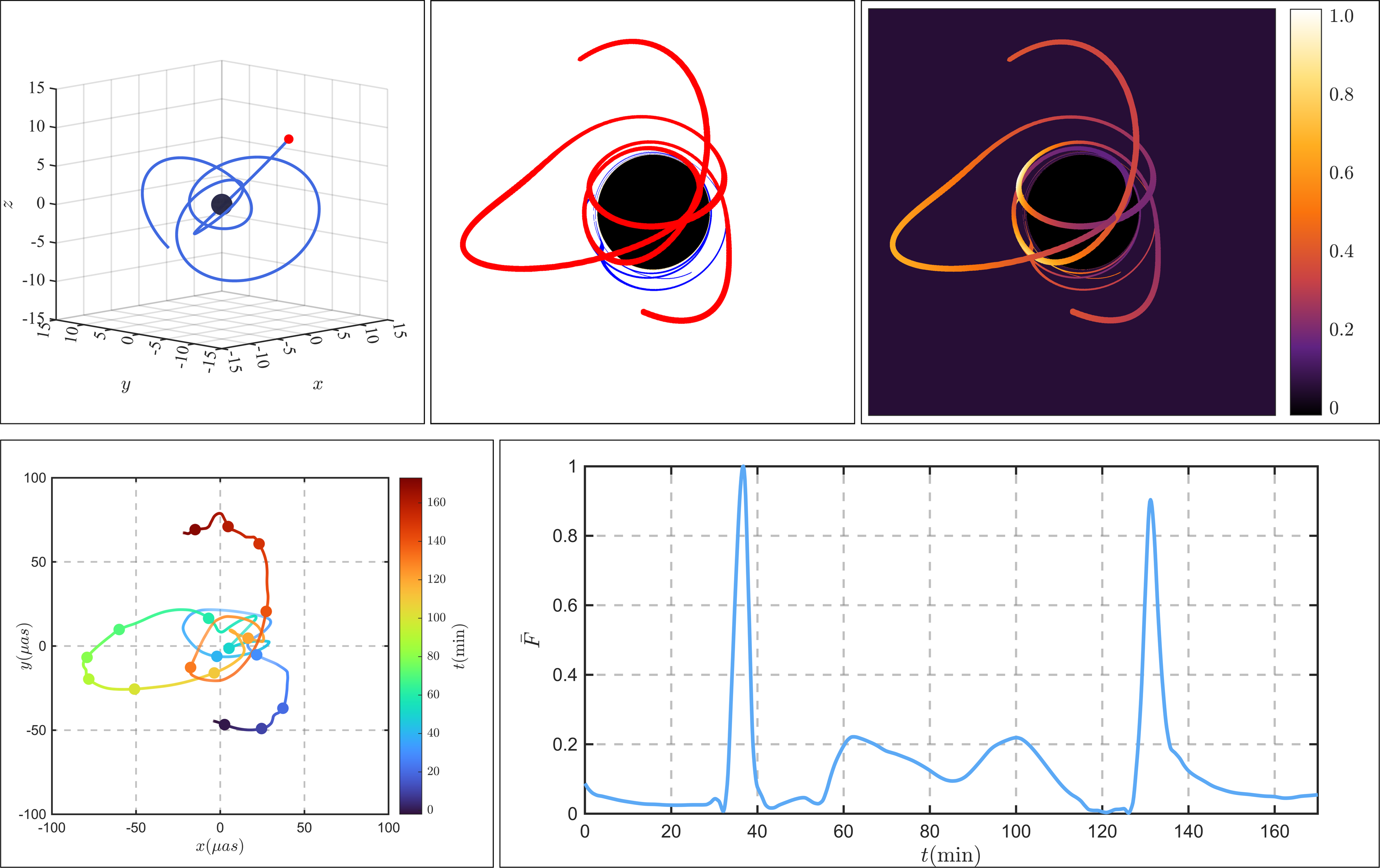}
  \centering
  \caption{The upper row of the illustration showcases the trajectory and image of a bounded hotspot. In contrast, the lower row visually represents the centroid motion and the corresponding light curve for this bounded orbit. The symbolism and interpretation of points and lines in this figure are consistent with those defined in Fig. \ref{fig:P}.}
  \label{fig:B}
\end{figure}

Next, we turn our attention to the images of bounded orbits, represented as $\M B$. As shown in Fig. \ref{fig:B} and similar to Fig. \ref{fig:P}, the top row depicts the trajectory and image of a bounded hotspot. The bottom row, on the other hand, shows the path of the emission centroid and its corresponding light curve for the same orbit. All these observations are conducted at an inclination angle of $\theta_o=25^\circ$.

For the bounded orbit we've chosen, as outlined in Table \ref{table:parameter of hotspot}, where $r_2<r_s<r_3$, the hotspot oscillates between the two turning points at $r_2$ and $r_3$. Fig. \ref{fig:B} depicts the hotspot's motion path as $r_s\to r_3\to r_2\to r_3\to r_2\to r_e<r_3$. Here, $r_e$ represents the endpoint of the trajectory in the radial direction. Due to the extensive trajectory of the hotspot's motion, both the primary (red) and secondary (blue) images on the observer's screen are relatively complex. Instead of detailing their intricate changes, we will focus on the motion of the centroid and causes of flares. For the emission centroid, we observe it moving back and forth within a finite radius, forming multiple rings. When the hotspot is far from the black hole, the brightness of the primary image far exceeds that of the secondary image, contributing more significantly to the emission centroid. Therefore, the motion of the centroid mainly reflects the trajectory of the primary image. As the hotspot approaches the vicinity of the black hole, the positions of the primary and secondary images are roughly symmetrically distributed about the black hole's shadow (for instance, when the primary image is above the shadow, the secondary image appears below it), and their brightness is approximately equal. This results in the emission centroid lingering within the black hole shadow. In terms of flux variation, we observe two distinct flares around the 35 min and 130 min marks. Additionally, there are two minor peaks noticeable around the 62 min and 100 min marks, connected by a concave line in between. The flares observed at the 35 min and 130 min points share a common origin, akin to the flare formation mechanism in deflecting orbits. These flares emerge when the hotspot navigates past the inner turning point, which is in close proximity to the black hole. Consequently, the strong gravitational force induces a dramatic shift in direction, and the hotspot's speed escalates rapidly post this turning point. Hence, we refer to these flares as TDBFs in our paper. The two minor peaks and the intervening trough occur between the two flares, corresponding to the hotspot's motion from $r_2 \to r_3 \to r_2$. Within this range, and not far from $r_3$, the radial motion direction of the hotspot forms an angle less than $90^\circ$ with the direction towards the observer. Hence, it can be essentially considered as moving towards the observer. Given that $r_3$ is relatively distant from the black hole, the gravitational influence is minor, and there is minimal change in the direction of velocity. Consequently, the primary factor affecting the flux is the magnitude of the velocity. In the $r_2 \to r_3 \to r_2$ process, the hotspot's velocity first increases, then decreases, and then increases and decreases again.

\subsection{$\M N=2$}

In this subsection, we will explore the features of images of orbits with $\M N=2$, as presented in Table \ref{table:parameter of hotspot}.

\subsubsection{Spherical orbits}

\begin{figure}[!ht]
  \centering
  \includegraphics[width=6in]{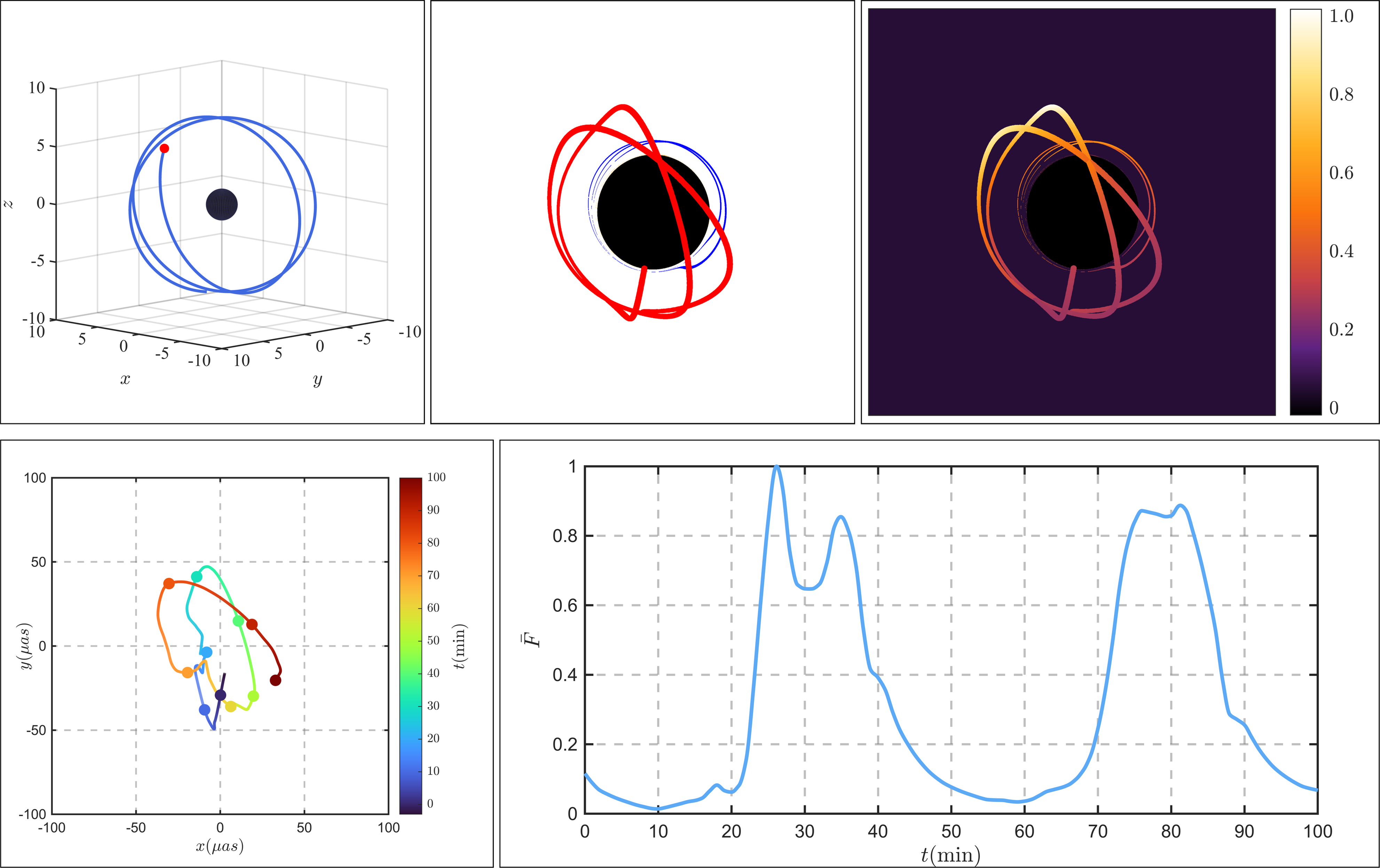}
  \centering
  \caption{The upper row of the illustration showcases the trajectory and image of a spherical hotspot. In contrast, the lower row visually represents the centroid motion and the corresponding light curve for this spherical orbit. The symbolism and interpretation of points and lines in this figure are consistent with those defined in Fig. \ref{fig:P}.}
  \label{fig:S}
\end{figure}

First, we focus on the spherical orbits, denoted as $\M S$. For these orbits, the hotspot is fixed at $r=r_s=8$. Therefore, the aspects that change in the trajectory are the polar and azimuthal coordinates. In Fig. \ref{fig:S}, we present an orbit with an azimuthal angle variation of $\Delta\varphi=4\pi$. The top row of the illustration displays the trajectory and image of a spherical hotspot, while the bottom row visually represents the centroid motion and the corresponding light curve for this spherical orbit. The symbols and interpretations of points and lines in this figure are consistent with those defined in Fig. \ref{fig:P}. The observation is conducted at an inclination angle of $\theta_o=25^\circ$.

From the middle image in the first row of Fig. \ref{fig:S}, it can be observed that the primary image of the hotspot takes the form of a precessing cap, while the secondary image appears on the opposite side of the primary image and has a fuller shape. From the intensity map on the right side of the first row in Fig. \ref{fig:S}, it can be seen that although the overall brightness of the secondary image is lower than that of the primary image, its brightness is still significant and should not be overlooked. Hence, the centroid of the hotspot image is influenced by both the primary and secondary images, resulting in an irregular trajectory of the centroid motion. The right plot in the second row of Fig. \ref{fig:S} displays a light curve with two prominent flares, each showing a split. These flares arise from the angular motion of the hotspot, which has a fixed radial position. When the hotspot moves to a certain angle, the Doppler blue shift peaks, leading to a flare in the primary image. This is swiftly followed by the blue-shift signal from the secondary image. The brief interval between these two signals gives the impression of a broad flare split. Indeed, when the time interval is reduced, the flare fully splits into two peaks, though this result has not been further explored in the paper. Given the spherical orbit's periodic angular motion, a flare corresponding to the first one around the 30 min mark is anticipated to appear around the 80 min mark, also showing a split. However, due to the precession in the hotspot's motion, the second flare's behavior doesn't entirely mirror that of the first. This type of flare is determined by both the angular Doppler blueshift and lensed Doppler blueshift. Therefore, we can conveniently refer to it as `Mixed Doppler Blueshift Flares', abbreviated as `MDBF'.

\subsubsection{Other orbits for $\M N=2$}

\begin{figure}[!ht]
  \centering
  \includegraphics[width=6in]{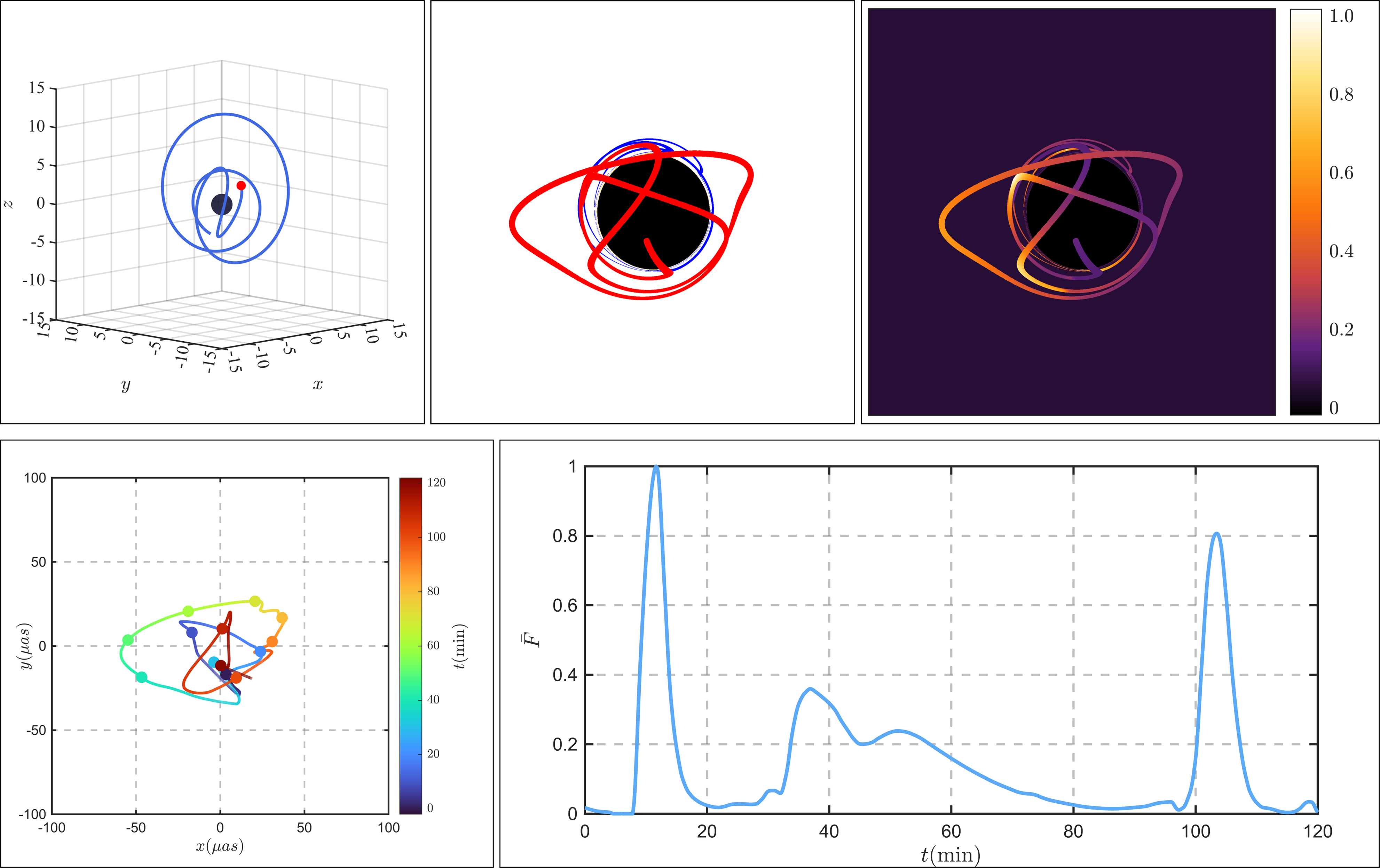}
  \centering
  \caption{In the illustration, the top row highlights the trajectory and image of a homoclinic hotspot. On the other hand, the bottom row graphically illustrates the centroid motion and the associated light curve for this same orbit. The symbols, points, and lines in this figure maintain consistency with those defined in Fig. \ref{fig:P}.}
  \label{fig:H}
\end{figure}

\begin{figure}[!ht]
  \centering
  \includegraphics[width=6in]{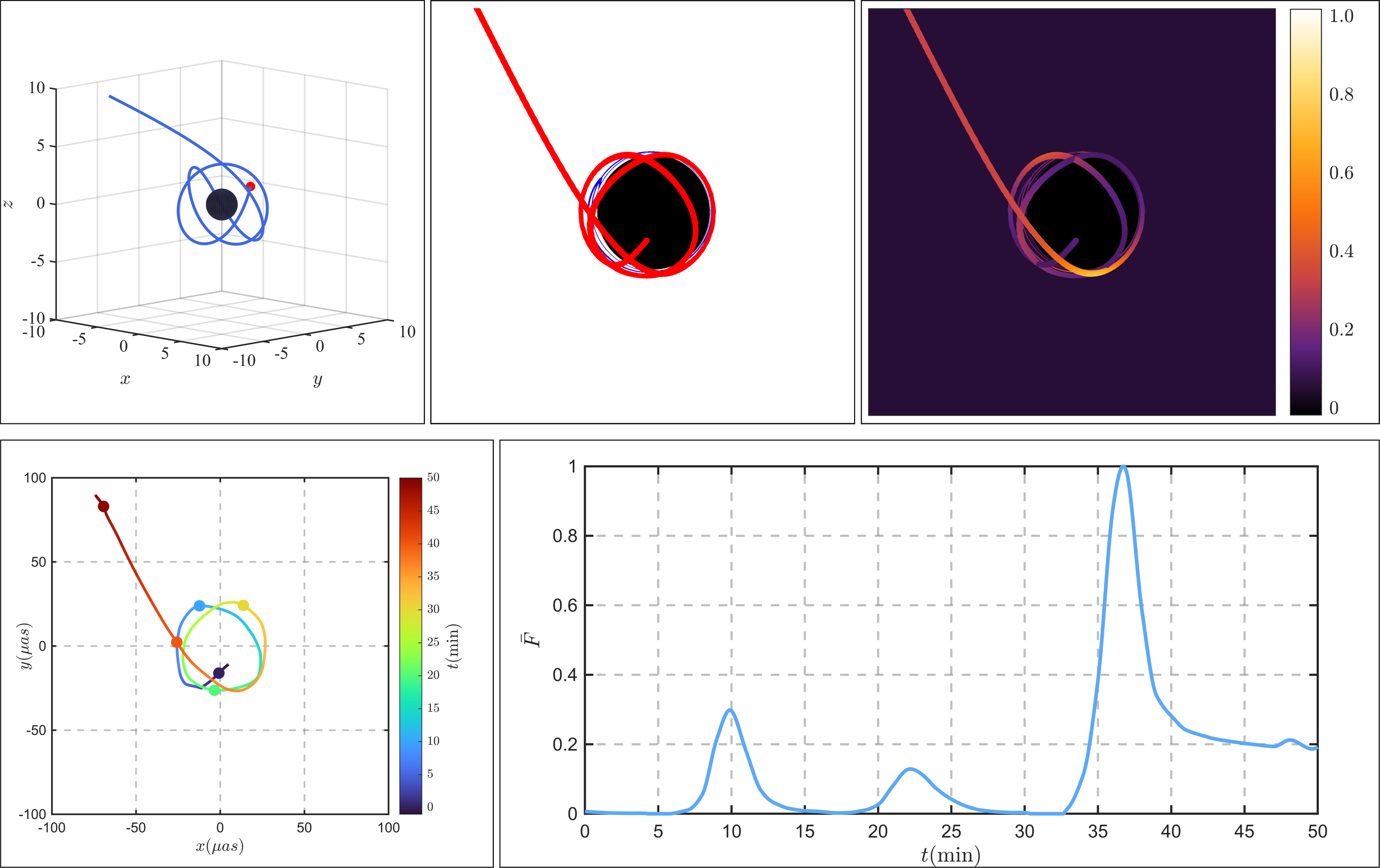}
  \centering
  \caption{The top row of the illustration demonstrates the trajectory and image of a whirling deflecting hotspot, whereas the bottom row provides a visual representation of the centroid motion and the associated light curve for this whirling deflecting orbit. The symbolism and interpretation of points and lines in this figure align with those established in Fig. \ref{fig:P}.}
  \label{fig:WD}
\end{figure}

\begin{figure}[!ht]
  \centering
  \includegraphics[width=6in]{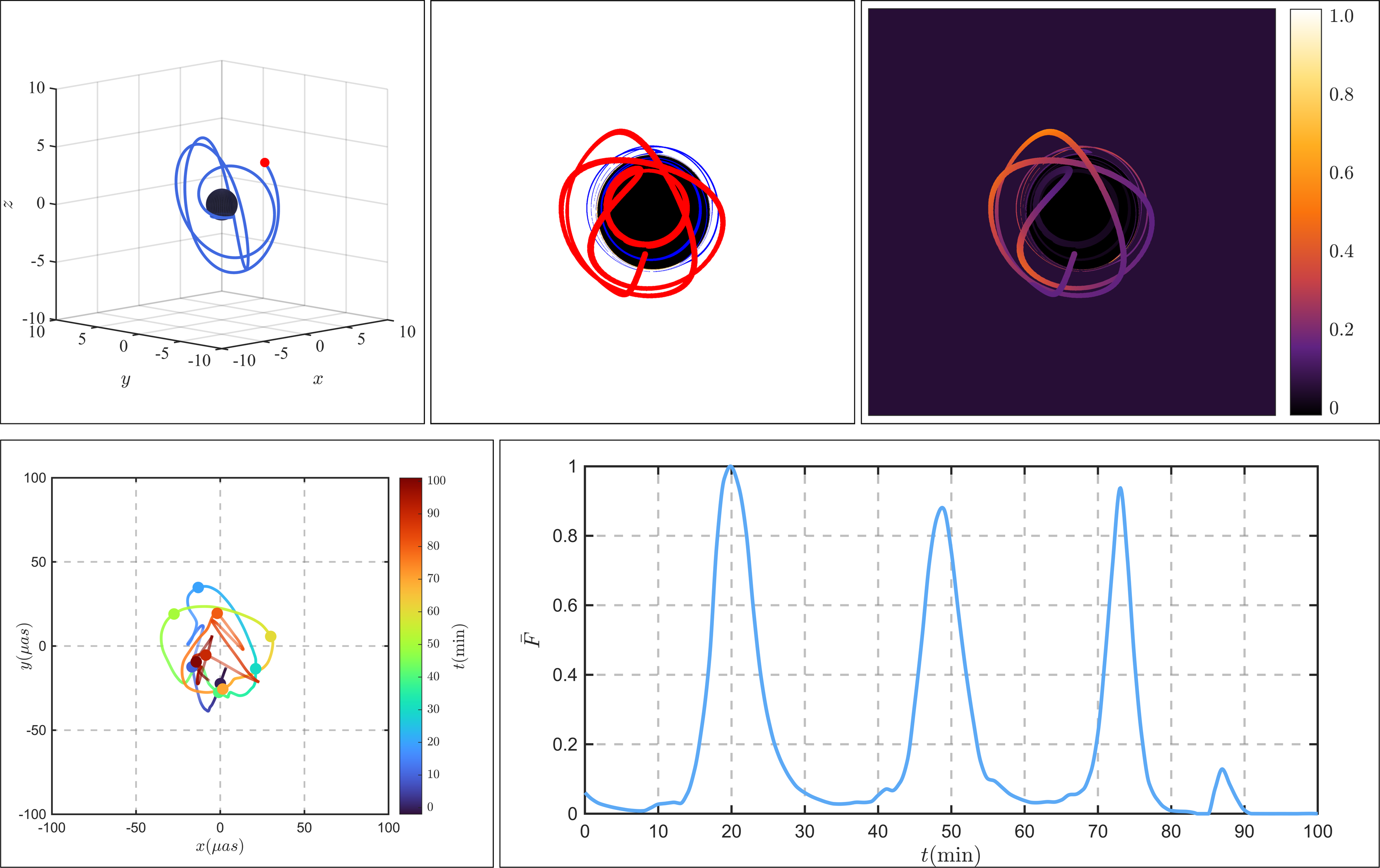}
  \centering
  \caption{In the illustration, the upper section presents the trajectory and image of a whirling trapped hotspot, while the lower section visually delineates the centroid motion and its corresponding light curve for this whirling trapped orbit. The symbology and interpretation of points and lines in this figure are in accordance with those specified in Fig. \ref{fig:P}.}
  \label{fig:WT}
\end{figure}

Next, we shift our focus to other orbits when $\M N=2$, namely, the homoclinic orbits $\M H$, the whirling deflecting orbits $\M{WD}$ and the whirling trapped orbits $\M{WT}$. We discuss the results of these three types of orbits collectively because they can be viewed as combinations of the previously mentioned orbits. The hotspot imaging and flares can also be explained accordingly using the conclusions drawn earlier.

\textbf{Homoclinic orbits} 

We know that in homoclinic orbits, a particle starts from an unstable spherical orbit, deflects at a turning point, and then returns to the original unstable spherical orbit. Clearly, the imaging characteristics during the unstable spherical orbit phase should resemble those of $\M S$. Considering the turning point is relatively distant from the black hole, its imaging characteristics should be similar to those near the outer turning point of $\M B$. In Fig. \ref{fig:H}, the imaging outcomes for homoclinic orbits are displayed, encompassing light intensity maps, graphs of centroid motion, and light curves. In comparing Fig. \ref{fig:H} to Fig. \ref{fig:S} and Fig. \ref{fig:B}, it's evident that aspects of both the light intensity maps and light curves from Fig. \ref{fig:S} and Fig. \ref{fig:B} are mirrored in Fig. \ref{fig:H}. The flares in Fig. \ref{fig:H} align with those in Fig. \ref{fig:S}, and the dual minor peaks along with the intermediate trough in Fig. \ref{fig:H} bear significant resemblance to their counterparts in Fig. \ref{fig:B}. However, there are some subtle discrepancies when Fig. \ref{fig:H} is juxtaposed with Fig. \ref{fig:S} and Fig. \ref{fig:B}. For example, the flare in Fig. \ref{fig:H} doesn't display any splitting, and the second minor peak in Fig. \ref{fig:H} is discernibly smaller than the first. Despite these minor variations, they don't alter the underlying cause of the flare, confirming that for $\M H$, these are also instances of MDBFs.

\textbf{Whirling deflecting orbits}

In Fig. \ref{fig:WD}, we present the results for a hotspot following a whirling deflecting orbit. This particular trajectory is selected because the hotspot originates from an unstable spherical orbit and then diverges toward infinity. As such, it can be considered a combination of spherical and deflecting orbits. Consequently, its imaging characteristics are a combination of the features found in spherical and deflecting orbits. Upon examining the light intensity plot in Fig. \ref{fig:WD}, it becomes evident that the hotspot mimics a spherical motion before it separates from the black hole. The intensity variation closely mirrors that of the spherical orbit represented in Fig. \ref{fig:S}. The phase of the hotspot's departure from the black hole aligns with the corresponding phase in Fig. \ref{fig:S}. Looking at the changes in the light curve, we notice two minor flares and one significantly larger flare. When these observations are compared with the imaging results from spherical and deflecting orbits, we find that the origins of the first two minor flares coincide with the results from the spherical orbit, characterized as MDBFs. The final and most substantial flare aligns with the results from the deflecting orbit, identified as a TDBF.

\textbf{Whirling trapped orbits}

Finally, we present the imaging results for whirling trapped orbits in Fig. \ref{fig:WT}. The whirling trapped orbit we chose for our work originates from a spherical orbit and then plunges into the black hole. Based on the characteristics of this orbit, we can consider it as a combination of spherical orbits and trapped orbits. The first stage is an approximate spherical orbit, and the second stage is a trapped orbit, which can be easily seen from the light intensity map in Fig. \ref{fig:WT}. We can see a total of three cap rings, and the last one has a noticeable deformation due to the hotspot deviating from the spherical orbit significantly. Eventually, it falls into the black hole shadow, consistent with the final stage of the light intensity map in Fig. \ref{fig:T}. From the light curve in Fig. \ref{fig:WT}, we can see three significant flares, corresponding to the three cap rings generated by the approximate spherical orbit. Therefore, the cause of these flares is consistent with the results from the spherical orbit, which are MDBFs. The last small flare matches the flare in Fig. \ref{fig:T}, so it is a LDBF.

\begin{figure}[!ht]
\vspace{-30pt}
\includegraphics[width=5.5in]{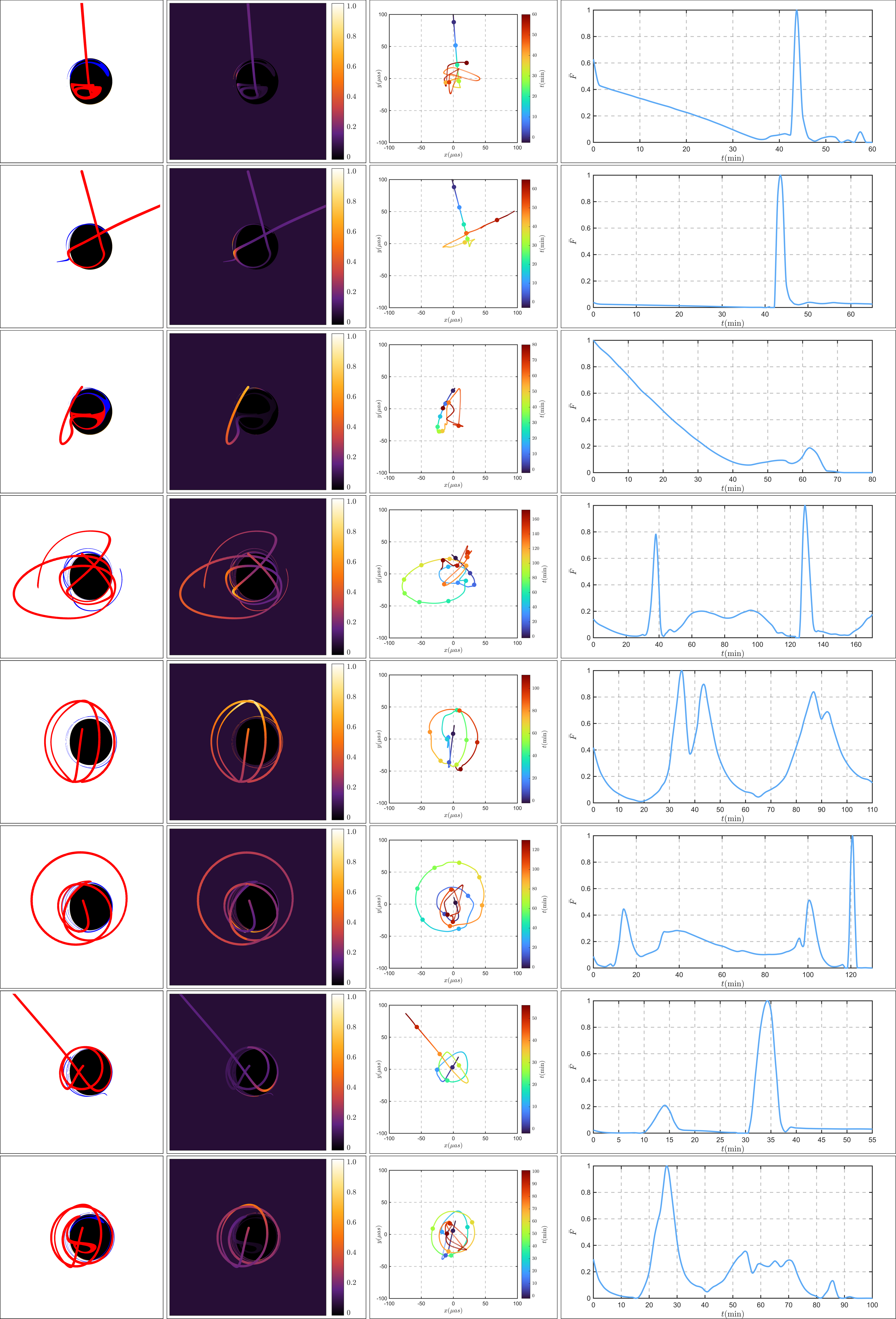}
  \centering
  \caption{The results at $\theta_o=80^\circ$ are presented, with the following order from top to bottom: plunging orbits $\M P$, deflecting orbits $\M D$, trapped orbits $\M T$, bounded orbits $\M B$, spherical orbits $\M S$, homoclinic orbits $\M H$, whirling deflecting orbits $\M{WD}$, and whirling trapped orbits $\M {WT}$. From left to right, they respectively represent the image of the hotspot, light intensity map, centroid motion, and light curve.}
  \label{fig:all80}
\end{figure}

\subsection{Results at $\theta_o=80^\circ$}

In this subsection, we also present the results for plunging orbits $\M P$, deflecting orbits $\M D$, trapped orbits $\M T$, bounded orbits $\M B$, spherical orbits $\M S$, homoclinic orbits $\M H$, whirling deflecting orbits $\M{WD}$, and whirling trapped orbits $\M {WT}$ at $\theta_o=80^\circ$ in 
Fig. \ref{fig:all80}.

We will not elaborate on the details of the image in Fig. \ref{fig:all80}. By comparing Fig. \ref{fig:all80} with the previous figures from Fig. \ref{fig:P} through to Fig. \ref{fig:WT}, we can see that the main features of the light intensity and flares of the hotspot image at an observation angle of $\theta_o=25^\circ$ still exist at an observation angle of $\theta_o=80^\circ$. This suggests that the characteristics of the hotspot image and flares we summarized earlier are not sensitive to the observation angle and can be observed at different angles.

\subsection{Einstein ring}

\begin{figure}[!ht]
  \centering
  \includegraphics[width=6in]{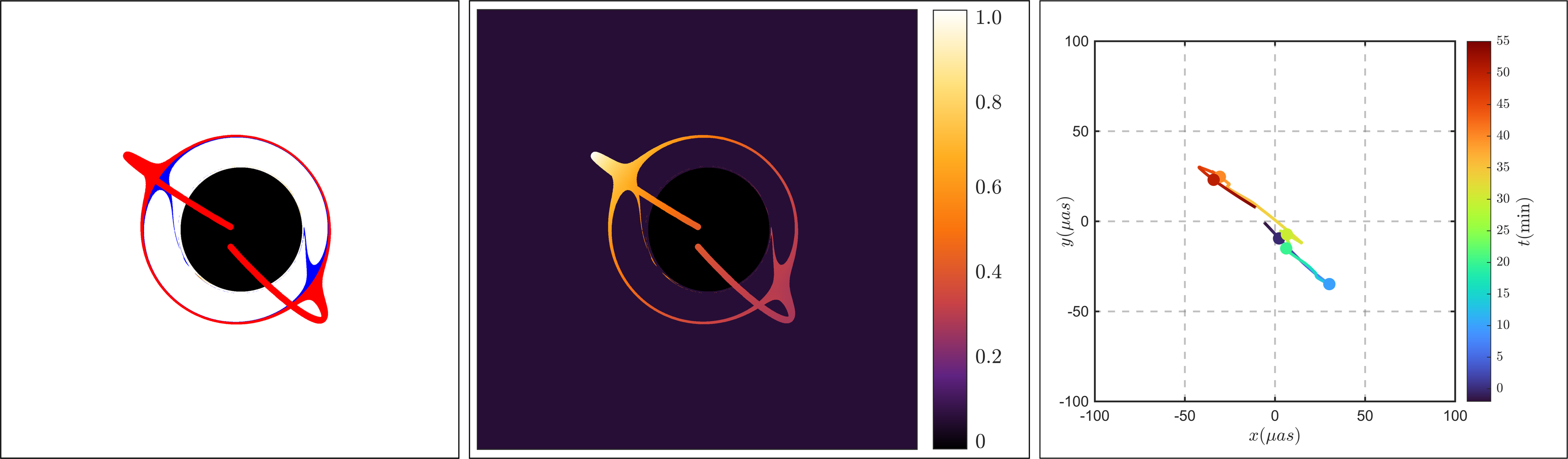}
  \centering
  \caption{The Einstein ring in the spherical case is shown. On the left is the image of the hotspot, in the middle is the light intensity map, and on the right is the plot of the centroid motion.}
  \label{fig:ring}
\end{figure}

Interestingly, during our investigation of the imaging of a hotspot moving in a spherical orbit, we discover that at specific instances, the hotspot aligns directly opposite the observer on the line connecting with the black hole, resulting in the appearance of an Einstein ring, as depicted in Fig. \ref{fig:ring}. However, as this situation is not the primary focus of our study, we will not delve into the details of this hotspot imaging, but merely present it here for reference.

\begin{table}[!ht]
  \centering
  \begin{tabular}{|c|c|c|c|c|c|c|c|c|}
  \hline
  Type   & $\M P$
  & $\M D$ & $\M T$ & $\M B$ & $\M S$ 
  & $\M H$ & $\M W \M{D}$ & $\M W \M{T}$
  \\ \hline
   Flare number 
  &$1$ &$1$ &$1$ &$2^{\ast}$
  &$2^{\ast }$ &$2^{\ast}$ &$3^{\ast}$ &$4^{\ast}$
  \\ \hline
  \multirow{2}*{Causes of flare}
  &\multirow{2}*{LDBF}
  &\multirow{2}*{TDBF}
  &\multirow{2}*{LDBF}
  &\multirow{2}*{TDBF}
  &\multirow{2}*{MDBFs}
  &\multirow{2}*{MDBFs}
  &{TDBF}
  &{LDBF}
  \\ 
    & & & & & &  &MDBFs &MDBFs
  \\
  \hline
  \end{tabular}
  \caption{The number of flares and their causes for various types of orbits, including plunging orbits $\M P$, deflecting orbits $\M D$, trapped orbits $\M T$, bounded orbits $\M B$, spherical orbits $\M S$, homoclinic orbits $\M H$, whirling deflecting orbits $\M{WD}$, and whirling trapped orbits $\M {WT}$ are discussed here. The `$\ast$' denote that these numbers represent the number of flares occurring within a particular period of interest in our study, rather than the total number of flares throughout the entire history of the hotspot. }
  \label{table:flare results of hotspot}
  \end{table}

\section{Summary}\label{sec5}

In our research, we explored the imaging of hotspots surrounding Kerr black holes. The hotspot, characterized as a luminous sphere with a radius of $b=0.25$, emits light uniformly and isotropically, independent of frequency. We incorporated the hotspot's movement along the geodesics outside a Kerr black hole into our analysis. Building on the conclusions of \cite{Compere:2021bkk}, we identified eight types of timelike geodesic orbits outside the Kerr black hole: plunging orbits $\M P$, deflecting orbits $\M D$, trapped orbits $\M T$, bounded orbits $\M B$, spherical orbits $\M S$, homoclinic orbits $\M H$, whirling deflecting orbits $\M{WD}$, and whirling trapped orbits $\M {WT}$. Importantly, our study extended beyond the equatorial plane. Using numerical backward ray tracing, we generated images of various orbital types. By calculating the fractional number, we differentiated between primary, secondary, and higher-order images and analyzed their distinct properties. We also tracked the temporal flux variation received by each pixel on the observer's image plane, allowing us to determine and analyze the motion of the centroid position across different orbits. Specifically, we computed the light curve of the hotspot image, observed flaring phenomena, and performed a detailed examination of these events. 

Specifically, we observed flare phenomena across all orbit types. We categorized flares into three types based on their origins: LDBFs, TDBFs, and MDBFs. The potential for each orbit to produce these flare types is summarized in Table \ref{table:flare results of hotspot}. Note that the numbers marked with `$\ast$' represent the count of flares within a specific period of interest in our study, not the total flare occurrences throughout the entire history of the hotspot. It's worth noting that for these orbit types, the majority of the motion characteristics remain consistent throughout the entire history of the hotspot. For bounded orbits $\M B$, the hotspot oscillates between two turning points along the radial direction; thus, analyzing one motion cycle suffices. For spherical orbits $\M S$, the hotspot continuously follows a spherical path, but only a single complete cycle in the polar coordinate $\theta$ needs to be analyzed. Similarly, for homoclinic orbits $\M H$, whirling deflecting orbits $\M{WD}$, and whirling trapped orbits $\M{WT}$, they all involve numerous near-spherical movements. Therefore, a clear analysis of a single cycle of near-spherical motion suffices.

We noted that LDBFs occur in orbits descending into the black hole, TDBFs in orbits with turning points near the black hole, and MDBFs in orbits traversing near the sphere. Furthermore, by comparing imaging results at $25^\circ$ and $80^\circ$, we determined that the flare phenomena we observed and the categorizations we established are not significantly influenced by the observation angle. Our findings provide further insights into the causes of flare formation and suggest that these common flare signals are likely detectable in astronomical observations.

We conclude our paper by outlining several future perspectives. First, our hotspot model has not yet accounted for the emission rates of the hotspot across various frequency bands—an aspect that necessitates further investigation. Second, considering the complex astronomical environment around black holes, the hotspot's trajectory might deviate from geodesics. Hence, it could be beneficial to delve deeper into the imaging characteristics of non-geodesic motion. Third, the existence of accretion disks and jets around the black hole could potentially influence the flare signals produced by the hotspot, a factor that deserves additional attention in future research. Lastly, it's important to note that the flares observed by the GRAVITY collaboration also carry polarization information. As such, the theoretical exploration of this polarization information is of substantial importance and warrants further study.

\section*{Acknowledgments}
We thank Yehui Hou for helpful discussions. The work is partly supported by NSFC Grant No. 12275004, 12205013 and 11873044. MG is also endorsed by "the Fundamental Research Funds for the Central Universities" with Grant No. 2021NTST13.

\appendix


\providecommand{\href}[2]{#2}\begingroup\raggedright\begin{thebibliography}{10}

\bibitem{EventHorizonTelescope:2019dse}
{\bfseries Event Horizon Telescope} Collaboration, K.~Akiyama {\em et~al.},
  ``{First M87 Event Horizon Telescope Results. I. The Shadow of the
  Supermassive Black Hole},''
  \href{http://dx.doi.org/10.3847/2041-8213/ab0ec7}{{\em Astrophys. J. Lett.}
  {\bfseries 875} (2019) L1}, \href{http://arxiv.org/abs/1906.11238}{{\ttfamily
  arXiv:1906.11238 [astro-ph.GA]}}.

\bibitem{EventHorizonTelescope:2022wkp}
{\bfseries Event Horizon Telescope} Collaboration, K.~Akiyama {\em et~al.},
  ``{First Sagittarius A* Event Horizon Telescope Results. I. The Shadow of the
  Supermassive Black Hole in the Center of the Milky Way},''
  \href{http://dx.doi.org/10.3847/2041-8213/ac6674}{{\em Astrophys. J. Lett.}
  {\bfseries 930} no.~2, (2022) L12},
  \href{http://arxiv.org/abs/2311.08680}{{\ttfamily arXiv:2311.08680
  [astro-ph.HE]}}.

\bibitem{GRAVITY:2018det}
{\bfseries GRAVITY} Collaboration, R.~Abuter {\em et~al.}, ``{Detection of
  orbital motions near the last stable circular orbit of the massive black hole
  SgrA},'' \href{http://dx.doi.org/10.1051/0004-6361/201834294}{{\em Astron.
  Astrophys.} {\bfseries 618} (2018) L10},
  \href{http://arxiv.org/abs/1810.12641}{{\ttfamily arXiv:1810.12641
  [astro-ph.GA]}}.

\bibitem{GRAVITY:2023avo}
{\bfseries GRAVITY} Collaboration, R.~Abuter {\em et~al.}, ``{Polarimetry and
  astrometry of NIR flares as event horizon scale, dynamical probes for the
  mass of Sgr A*},'' \href{http://dx.doi.org/10.1051/0004-6361/202347416}{{\em
  Astron. Astrophys.} {\bfseries 677} (2023) L10},
  \href{http://arxiv.org/abs/2307.11821}{{\ttfamily arXiv:2307.11821
  [astro-ph.GA]}}.

\bibitem{Gyulchev:2020cvo}
G.~Gyulchev, J.~Kunz, P.~Nedkova, T.~Vetsov, and S.~Yazadjiev, ``{Observational
  signatures of strongly naked singularities: image of the thin accretion
  disk},'' \href{http://dx.doi.org/10.1140/epjc/s10052-020-08575-7}{{\em Eur.
  Phys. J. C} {\bfseries 80} no.~11, (2020) 1017},
  \href{http://arxiv.org/abs/2003.06943}{{\ttfamily arXiv:2003.06943 [gr-qc]}}.

\bibitem{Guerrero:2021ues}
M.~Guerrero, G.~J. Olmo, D.~Rubiera-Garcia, and D.~S.-C. G\'omez, ``{Shadows
  and optical appearance of black bounces illuminated by a thin accretion
  disk},'' \href{http://dx.doi.org/10.1088/1475-7516/2021/08/036}{{\em JCAP}
  {\bfseries 08} (2021) 036}, \href{http://arxiv.org/abs/2105.15073}{{\ttfamily
  arXiv:2105.15073 [gr-qc]}}.

\bibitem{Hou:2022eev}
Y.~Hou, Z.~Zhang, H.~Yan, M.~Guo, and B.~Chen, ``{Image of a Kerr-Melvin black
  hole with a thin accretion disk},''
  \href{http://dx.doi.org/10.1103/PhysRevD.106.064058}{{\em Phys. Rev. D}
  {\bfseries 106} no.~6, (2022) 064058},
  \href{http://arxiv.org/abs/2206.13744}{{\ttfamily arXiv:2206.13744 [gr-qc]}}.

\bibitem{Qin:2023nog}
X.~Qin, S.~Chen, Z.~Zhang, and J.~Jing, ``{Polarized image of a rotating black
  hole surrounded by a cold dark matter halo},''
  \href{http://dx.doi.org/10.1140/epjc/s10052-023-11300-9}{{\em Eur. Phys. J.
  C} {\bfseries 83} no.~2, (2023) 159},
  \href{http://arxiv.org/abs/2301.01551}{{\ttfamily arXiv:2301.01551 [gr-qc]}}.

\bibitem{Meng:2023htc}
Y.~Meng, X.-M. Kuang, X.-J. Wang, B.~Wang, and J.-P. Wu, ``{Images from disk
  and spherical accretions of hairy Schwarzschild black holes},''
  \href{http://dx.doi.org/10.1103/PhysRevD.108.064013}{{\em Phys. Rev. D}
  {\bfseries 108} no.~6, (2023) 064013},
  \href{http://arxiv.org/abs/2306.10459}{{\ttfamily arXiv:2306.10459 [gr-qc]}}.

\bibitem{Guo:2023zwy}
S.~Guo, Y.-X. Huang, and G.-P. Li, ``{Optical appearance of the Schwarzschild
  black hole in the string cloud context*},''
  \href{http://dx.doi.org/10.1088/1674-1137/accad5}{{\em Chin. Phys. C}
  {\bfseries 47} no.~6, (2023) 065105},
  \href{http://arxiv.org/abs/2305.00007}{{\ttfamily arXiv:2305.00007 [gr-qc]}}.

\bibitem{Zhang:2024lsf}
Z.~Zhang, Y.~Hou, M.~Guo, and B.~Chen, ``{Imaging thick accretion disks and
  jets surrounding black holes},''
  \href{http://arxiv.org/abs/2401.14794}{{\ttfamily arXiv:2401.14794
  [astro-ph.HE]}}.

\bibitem{Ripperda:2021zpn}
B.~Ripperda, M.~Liska, K.~Chatterjee, G.~Musoke, A.~A. Philippov, S.~B.
  Markoff, A.~Tchekhovskoy, and Z.~Younsi, ``{Black Hole Flares: Ejection of
  Accreted Magnetic Flux through 3D Plasmoid-mediated Reconnection},''
  \href{http://dx.doi.org/10.3847/2041-8213/ac46a1}{{\em Astrophys. J. Lett.}
  {\bfseries 924} no.~2, (2022) L32},
  \href{http://arxiv.org/abs/2109.15115}{{\ttfamily arXiv:2109.15115
  [astro-ph.HE]}}.

\bibitem{GRAVITY:2021hxs}
{\bfseries GRAVITY} Collaboration, R.~Abuter {\em et~al.}, ``{Constraining
  particle acceleration in Sgr A\ensuremath{\star} with simultaneous GRAVITY,
  Spitzer, NuSTAR, and Chandra observations},''
  \href{http://dx.doi.org/10.1051/0004-6361/202140981}{{\em Astron. Astrophys.}
  {\bfseries 654} (2021) A22},
  \href{http://arxiv.org/abs/2107.01096}{{\ttfamily arXiv:2107.01096
  [astro-ph.HE]}}.

\bibitem{Dexter:2020cuv}
J.~Dexter {\em et~al.}, ``{Sgr A* near-infrared flares from reconnection events
  in a magnetically arrested disc},''
  \href{http://dx.doi.org/10.1093/mnras/staa2288}{{\em Mon. Not. Roy. Astron.
  Soc.} {\bfseries 497} no.~4, (2020) 4999--5007},
  \href{http://arxiv.org/abs/2006.03657}{{\ttfamily arXiv:2006.03657
  [astro-ph.HE]}}.

\bibitem{Porth:2020txf}
O.~Porth, Y.~Mizuno, Z.~Younsi, and C.~M. Fromm, ``{Flares in the Galactic
  Centre \textendash{} I. Orbiting flux tubes in magnetically arrested black
  hole accretion discs},'' \href{http://dx.doi.org/10.1093/mnras/stab163}{{\em
  Mon. Not. Roy. Astron. Soc.} {\bfseries 502} no.~2, (2021) 2023--2032},
  \href{http://arxiv.org/abs/2006.03658}{{\ttfamily arXiv:2006.03658
  [astro-ph.HE]}}.

\bibitem{Ripperda:2020bpz}
B.~Ripperda, F.~Bacchini, and A.~Philippov, ``{Magnetic Reconnection and Hot
  Spot Formation in Black Hole Accretion Disks},''
  \href{http://dx.doi.org/10.3847/1538-4357/ababab}{{\em Astrophys. J.}
  {\bfseries 900} no.~2, (2020) 100},
  \href{http://arxiv.org/abs/2003.04330}{{\ttfamily arXiv:2003.04330
  [astro-ph.HE]}}.

\bibitem{Scepi:2021xgs}
N.~Scepi, J.~Dexter, and M.~C. Begelman, ``{Sgr A* X-ray flares from
  non-thermal particle acceleration in a magnetically arrested disc},''
  \href{http://dx.doi.org/10.1093/mnras/stac337}{{\em Mon. Not. Roy. Astron.
  Soc.} {\bfseries 511} no.~3, (2022) 3536--3547},
  \href{http://arxiv.org/abs/2107.08056}{{\ttfamily arXiv:2107.08056
  [astro-ph.HE]}}.

\bibitem{Jia:2023iup}
H.~Jia, B.~Ripperda, E.~Quataert, C.~J. White, K.~Chatterjee, A.~Philippov, and
  M.~Liska, ``{Millimeter observational signatures of flares in magnetically
  arrested black hole accretion models},''
  \href{http://dx.doi.org/10.1093/mnras/stad2935}{{\em Mon. Not. Roy. Astron.
  Soc.} {\bfseries 526} no.~2, (2023) 2924--2941},
  \href{http://arxiv.org/abs/2301.09014}{{\ttfamily arXiv:2301.09014
  [astro-ph.HE]}}.

\bibitem{Yuan:2003dc}
F.~Yuan, E.~Quataert, and R.~Narayan, ``{Nonthermal electrons in radiatively
  inefficient accretion flow models of Sagittarius A*},''
  \href{http://dx.doi.org/10.1086/378716}{{\em Astrophys. J.} {\bfseries 598}
  (2003) 301--312}, \href{http://arxiv.org/abs/astro-ph/0304125}{{\ttfamily
  arXiv:astro-ph/0304125}}.

\bibitem{Younsi:2015xna}
Z.~Younsi and K.~Wu, ``{Variations in emission from episodic plasmoid ejecta
  around black holes},'' \href{http://dx.doi.org/10.1093/mnras/stv2203}{{\em
  Mon. Not. Roy. Astron. Soc.} {\bfseries 454} no.~3, (2015) 3283--3298},
  \href{http://arxiv.org/abs/1510.01700}{{\ttfamily arXiv:1510.01700
  [astro-ph.HE]}}.

\bibitem{ball2021plasmoid}
D.~Ball, F.~{\"O}zel, P.~Christian, C.-K. Chan, and D.~Psaltis, ``A plasmoid
  model for the sgr a* flares observed with gravity and chandra,'' {\em The
  Astrophysical Journal} {\bfseries 917} no.~1, (2021) 8.

\bibitem{Aimar:2023kzj}
N.~Aimar, A.~Dmytriiev, F.~H. Vincent, I.~E. Mellah, T.~Paumard, G.~Perrin, and
  A.~Zech, ``{Magnetic reconnection plasmoid model for Sagittarius A*
  flares},'' \href{http://dx.doi.org/10.1051/0004-6361/202244936}{{\em Astron.
  Astrophys.} {\bfseries 672} (2023) A62},
  \href{http://arxiv.org/abs/2301.11874}{{\ttfamily arXiv:2301.11874
  [astro-ph.HE]}}.

\bibitem{Rowan:2017cao}
M.~E. Rowan, L.~Sironi, and R.~Narayan, ``{Electron and Proton Heating in
  Transrelativistic Magnetic Reconnection},''
  \href{http://dx.doi.org/10.3847/1538-4357/aa9380}{{\em Astrophys. J.}
  {\bfseries 850} no.~1, (2017) 29},
  \href{http://arxiv.org/abs/1708.04627}{{\ttfamily arXiv:1708.04627
  [astro-ph.HE]}}.

\bibitem{Levinson:2018arx}
A.~Levinson and B.~Cerutti, ``{Particle-in-cell simulations of pair discharges
  in a starved magnetosphere of a Kerr black hole},''
  \href{http://dx.doi.org/10.1051/0004-6361/201832915}{{\em Astron. Astrophys.}
  {\bfseries 616} (2018) A184},
  \href{http://arxiv.org/abs/1803.04427}{{\ttfamily arXiv:1803.04427
  [astro-ph.HE]}}.

\bibitem{Galishnikova:2022mjg}
A.~Galishnikova, A.~Philippov, E.~Quataert, F.~Bacchini, K.~Parfrey, and
  B.~Ripperda, ``{Collisionless Accretion onto Black Holes: Dynamics and
  Flares},'' \href{http://dx.doi.org/10.1103/PhysRevLett.130.115201}{{\em Phys.
  Rev. Lett.} {\bfseries 130} no.~11, (2023) 115201},
  \href{http://arxiv.org/abs/2212.02583}{{\ttfamily arXiv:2212.02583
  [astro-ph.HE]}}.

\bibitem{ElMellah:2023sun}
I.~El~Mellah, B.~Cerutti, and B.~Crinquand, ``{Reconnection-driven flares in 3D
  black hole magnetospheres - A scenario for hot spots around Sagittarius
  A*},'' \href{http://dx.doi.org/10.1051/0004-6361/202346781}{{\em Astron.
  Astrophys.} {\bfseries 677} (2023) A67},
  \href{http://arxiv.org/abs/2305.01689}{{\ttfamily arXiv:2305.01689
  [astro-ph.HE]}}.

\bibitem{Zhdankin:2023wch}
V.~Zhdankin, B.~Ripperda, and A.~A. Philippov, ``{Particle acceleration by
  magnetic Rayleigh-Taylor instability: Mechanism for flares in black hole
  accretion flows},''
  \href{http://dx.doi.org/10.1103/PhysRevResearch.5.043023}{{\em Phys. Rev.
  Res.} {\bfseries 5} no.~4, (2023) 043023},
  \href{http://arxiv.org/abs/2302.05276}{{\ttfamily arXiv:2302.05276
  [astro-ph.HE]}}.

\bibitem{Broderick:2005my}
A.~E. Broderick and A.~Loeb, ``{Imaging bright spots in the accretion flow near
  the black hole horizon of Sgr A*},''
  \href{http://dx.doi.org/10.1111/j.1365-2966.2005.09458.x}{{\em Mon. Not. Roy.
  Astron. Soc.} {\bfseries 363} (2005) 353--362},
  \href{http://arxiv.org/abs/astro-ph/0506433}{{\ttfamily
  arXiv:astro-ph/0506433}}.

\bibitem{Meyer:2006fd}
L.~Meyer, A.~Eckart, R.~Schoedel, W.~J. Duschl, K.~Muzic, M.~Dovciak, and
  V.~Karas, ``{Near-infrared polarimetry setting constraints on the orbiting
  spot model for Sgr A* flares},''
  \href{http://dx.doi.org/10.1051/0004-6361:20065925}{{\em Astron. Astrophys.}
  {\bfseries 460} (2006) 15},
  \href{http://arxiv.org/abs/astro-ph/0610104}{{\ttfamily
  arXiv:astro-ph/0610104}}.

\bibitem{Trippe:2006jy}
S.~Trippe, T.~Paumard, T.~Ott, S.~Gillessen, F.~Eisenhauer, F.~Martins, and
  R.~Genzel, ``{A polarised infrared flare from Sagittarius A* and the
  signatures of orbiting plasma hotspots},''
  \href{http://dx.doi.org/10.1111/j.1365-2966.2006.11338.x}{{\em Mon. Not. Roy.
  Astron. Soc.} {\bfseries 375} (2007) 764--772},
  \href{http://arxiv.org/abs/astro-ph/0611737}{{\ttfamily
  arXiv:astro-ph/0611737}}.

\bibitem{Hamaus:2008yw}
N.~Hamaus, T.~Paumard, T.~Muller, S.~Gillessen, F.~Eisenhauer, S.~Trippe, and
  R.~Genzel, ``{Prospects for testing the nature of Sgr A*'s NIR flares on the
  basis of current VLT- and future VLTI-observations},''
  \href{http://dx.doi.org/10.1088/0004-637X/692/1/902}{{\em Astrophys. J.}
  {\bfseries 692} (2009) 902--916},
  \href{http://arxiv.org/abs/0810.4947}{{\ttfamily arXiv:0810.4947
  [astro-ph]}}.

\bibitem{Zamaninasab:2009df}
M.~Zamaninasab {\em et~al.}, ``{Near infrared flares of Sagittarius A*:
  Importance of near infrared polarimetry},''
  \href{http://dx.doi.org/10.1051/0004-6361/200912473}{{\em Astron. Astrophys.}
  {\bfseries 510} (2010) A3}, \href{http://arxiv.org/abs/0911.4659}{{\ttfamily
  arXiv:0911.4659 [astro-ph.GA]}}.

\bibitem{Dokuchaev:2020rye}
V.~I. Dokuchaev and N.~O. Nazarova, ``{Modeling the motion of a bright spot in
  jets from black holes M87* and SgrA*},''
  \href{http://dx.doi.org/10.1007/s10714-021-02854-8}{{\em Gen. Rel. Grav.}
  {\bfseries 53} no.~8, (2021) 83},
  \href{http://arxiv.org/abs/2010.01885}{{\ttfamily arXiv:2010.01885
  [astro-ph.HE]}}.

\bibitem{Rosa:2023qcv}
J.~a.~L. Rosa, C.~F.~B. Macedo, and D.~Rubiera-Garcia, ``{Imaging compact boson
  stars with hot spots and thin accretion disks},''
  \href{http://dx.doi.org/10.1103/PhysRevD.108.044021}{{\em Phys. Rev. D}
  {\bfseries 108} no.~4, (2023) 044021},
  \href{http://arxiv.org/abs/2303.17296}{{\ttfamily arXiv:2303.17296 [gr-qc]}}.

\bibitem{Rosa:2024bqv}
J.~a.~L. Rosa, D.~S.~J. Cordeiro, C.~F.~B. Macedo, and F.~S.~N. Lobo,
  ``{Observational imprints of gravastars from accretion disks and
  hot-spots},'' \href{http://arxiv.org/abs/2401.07766}{{\ttfamily
  arXiv:2401.07766 [gr-qc]}}.

\bibitem{Chen:2023qic}
Y.~Chen, P.~Wang, and H.~Yang, ``{Interferometric Signatures of Black Holes
  with Multiple Photon Spheres},''
  \href{http://arxiv.org/abs/2312.10304}{{\ttfamily arXiv:2312.10304 [gr-qc]}}.

\bibitem{Chen:2024ilc}
Y.~Chen, P.~Wang, and H.~Yang, ``{Observations of Orbiting Hot Spots around
  Scalarized Reissner-Nordstr\"om Black Holes},''
  \href{http://arxiv.org/abs/2401.10905}{{\ttfamily arXiv:2401.10905 [gr-qc]}}.

\bibitem{Matsumoto:2020wul}
T.~Matsumoto, C.-H. Chan, and T.~Piran, ``{The origin of hotspots around Sgr
  A*: Orbital or pattern motion?},''
  \href{http://dx.doi.org/10.1093/mnras/staa2095}{{\em Mon. Not. Roy. Astron.
  Soc.} {\bfseries 497} no.~2, (2020) 2385--2392},
  \href{http://arxiv.org/abs/2004.13029}{{\ttfamily arXiv:2004.13029
  [astro-ph.HE]}}.

\bibitem{Wielgus:2022heh}
M.~Wielgus, M.~Moscibrodzka, J.~Vos, Z.~Gelles, I.~Marti-Vidal, J.~Farah,
  N.~Marchili, C.~Goddi, and H.~Messias, ``{Orbital motion near Sagittarius A*
  - Constraints from polarimetric ALMA observations},''
  \href{http://dx.doi.org/10.1051/0004-6361/202244493}{{\em Astron. Astrophys.}
  {\bfseries 665} (2022) L6}, \href{http://arxiv.org/abs/2209.09926}{{\ttfamily
  arXiv:2209.09926 [astro-ph.HE]}}.

\bibitem{vonFellenberg:2023hit}
S.~D. von Fellenberg {\em et~al.}, ``{General relativistic effects and the
  near-infrared and X-ray variability of Sgr A* I},''
  \href{http://dx.doi.org/10.1051/0004-6361/202245575}{{\em Astron. Astrophys.}
  {\bfseries 669} (2023) L17},
  \href{http://arxiv.org/abs/2301.02558}{{\ttfamily arXiv:2301.02558
  [astro-ph.HE]}}.

\bibitem{GRAVITY:2020lpa}
{\bfseries GRAVITY} Collaboration, M.~Baub\"ock {\em et~al.}, ``{Modeling the
  orbital motion of Sgr A*\textquoteright{}s near-infrared flares},''
  \href{http://dx.doi.org/10.1051/0004-6361/201937233}{{\em Astron. Astrophys.}
  {\bfseries 635} (2020) A143},
  \href{http://arxiv.org/abs/2002.08374}{{\ttfamily arXiv:2002.08374
  [astro-ph.HE]}}.

\bibitem{Compere:2021bkk}
G.~Comp\`ere, Y.~Liu, and J.~Long, ``{Classification of radial Kerr geodesic
  motion},'' \href{http://dx.doi.org/10.1103/PhysRevD.105.024075}{{\em Phys.
  Rev. D} {\bfseries 105} no.~2, (2022) 024075},
  \href{http://arxiv.org/abs/2106.03141}{{\ttfamily arXiv:2106.03141 [gr-qc]}}.

\bibitem{Bardeen:1973tla}
J.~M. Bardeen, ``{Timelike and null geodesics in the Kerr metric},'' {\em
  Proceedings, Ecole d'Et\'e de Physique Th\'eorique: Les Astres Occlus : Les
  Houches, France, August, 1972, 215-240} (1973) 215--240.

\bibitem{Hu:2020usx}
Z.~Hu, Z.~Zhong, P.-C. Li, M.~Guo, and B.~Chen, ``{QED effect on a black hole
  shadow},'' \href{http://dx.doi.org/10.1103/PhysRevD.103.044057}{{\em Phys.
  Rev. D} {\bfseries 103} no.~4, (2021) 044057},
  \href{http://arxiv.org/abs/2012.07022}{{\ttfamily arXiv:2012.07022 [gr-qc]}}.

\bibitem{Lindquist:1966igj}
R.~W. Lindquist, ``{Relativistic transport theory},''
  \href{http://dx.doi.org/10.1016/0003-4916(66)90207-7}{{\em Annals Phys.}
  {\bfseries 37} no.~3, (1966) 487--518}.

\bibitem{Zhang:2023cuw}
Z.~Zhang, Y.~Hou, Z.~Hu, M.~Guo, and B.~Chen, ``{Polarized images of charged
  particles in vortical motions around a magnetized Kerr black hole},''
  \href{http://arxiv.org/abs/2304.03642}{{\ttfamily arXiv:2304.03642 [gr-qc]}}.

\bibitem{rybicki1991radiative}
G.~B. Rybicki and A.~P. Lightman, {\em Radiative processes in astrophysics}.
\newblock John Wiley \& Sons, 1991.

\end{thebibliography}\endgroup
\end{document}